\documentclass[12pt,draftcls,onecolumn]{IEEEtran}

\usepackage{amsmath,amssymb,graphicx,dsfont,cite,color}
\usepackage{enumerate}
\usepackage{bm, upgreek}
\usepackage{url}
\usepackage{arydshln}

\newtheorem{theorem}{Theorem}
\newtheorem{lemma}{Lemma}
\newtheorem{prop}{Proposition}

\newtheorem{defn}{Definition}
\newtheorem{remark}{Remark}
\newtheorem{example}{Example}

\begin{document}
\title{A Principled Approximation Framework for Optimal Control of Semi-Markov Jump Linear Systems}
\author{Saeid Jafari and Ketan Savla
\thanks{The authors are with the Sonny Astani Department of Civil and Environmental Engineering at the University of Southern California, Los Angeles, CA, 90089, USA. Email: {\tt\small\{sjafari,ksavla\}@usc.edu}.}
}
\maketitle
\bibliographystyle{IEEEtran}
\begin{abstract}

We consider continuous-time, finite-horizon, optimal quadratic control of semi-Markov jump linear systems (S-MJLS), and develop principled approximations through Markov-like representations for the holding-time distributions. 
We adopt a phase-type approximation for holding times, which is known to be consistent, and translates a S-MJLS into a specific MJLS with partially observable modes (MJLSPOM), where the modes in a cluster have the same dynamic, the same cost weighting matrices and the same control policy. For a general MJLSPOM, we give necessary and sufficient conditions for optimal (switched) linear controllers. When specialized to our particular MJLSPOM, we additionally establish the existence of optimal linear controller, as well as its optimality within the class of general controllers satisfying standard smoothness conditions. 
The known equivalence between phase-type distributions and positive linear systems allows to leverage existing modeling tools, but possibly with large computational costs. Motivated by this, we propose matrix exponential approximation of holding times, resulting in \emph{pseudo}-MJLSPOM representation, i.e., where the transition rates could be negative. Such a representation is of relatively low order, and maintains the same optimality conditions as for the MJLSPOM representation, but could violate non-negativity of holding-time density functions. A two-step procedure consisting of a local pulling-up modification and a filtering technique is constructed to enforce non-negativity. 

\end{abstract}
\IEEEpeerreviewmaketitle
\section{Introduction}

In many engineering applications, systems may experience random abrupt variations in their parameters and structure that change the system's dynamic and operating condition. Examples include power systems with randomly varying loads, systems whose operating condition depends on random phenomena such as wind speed and solar irradiance,  networked control systems with sudden changes due to random variations in the network topology, and avionic systems in the presence of electromagnetic disturbances from both natural and man-made sources \cite{Loparo90, Abdollahi2011, Shooman1994}. For other applications, see  \cite[\S1.3]{Mariton90}, \cite[\S1.2]{Costa13}, and references therein.

Due to tractability of linear models for control and optimization purposes, systems subject to random changes are often modeled by \emph{stochastic jump linear systems}, consisting of a finite number of linear models where switching among them is governed by an exogenous random process. 
Modeling of the jump process is carried out by fitting a suitable probability model to historical data on the sequence of jump times and waiting times in each mode. 
%
The \emph{homogeneous Markov chain}, due to its mathematical tractability, is the most commonly used stochastic model for a jump process, for the purpose of analysis and control design. However, the memoryless property forces the holding time in each mode to be exponentially distributed, while many features of real systems are not memoryless. 

The \emph{semi-Markov process} is a generalization of the Markov chain, in which the distribution of the time the process spends in any mode before jumping to another is allowed to be non-exponential. In many applications, the semi-Markov process is a natural stochastic model to describe a random process with a discrete state space. For example, the semi-Markov process is  a suitable model to describe the operating characteristics of power plants, and to assess reliability of power systems \cite{Perman97}. Similarly, for optimization and reliability analysis of wind turbines, the wind speed process is often modeled by a semi-Markov process, as it more accurately reproduces the statistical properties of wind speed data compared to a Markov process \cite{DAmico13, DAmico15}.

However, mathematical analysis of controlled dynamical systems consisting of a non-Markovian jump process is often difficult. In order to arrive at a tractable method for analysis and design, one approach is to transform the non-Markovian process into a finite-state homogeneous Markov model, by including sufficient supplementary state variables to model some part of the process history \cite[\S2.3.7]{Bolch06}. In reliability theory, a commonly-used approach to model non-exponential life-time distributions\footnote{ By \emph{life-time distributions}, we mean any continuous distribution with support on the non-negative real numbers.} is approximation by a class of distributions called \emph{phase-type distribution} (or \emph{PH distribution}, for short) \cite{Bolch06}.  The PH distribution is a generalization of the exponential distribution, and is defined as the distribution of the time to enter an absorbing state from a set of transient states in a finite-state Markov chain. The PH distributions are dense (in the sense of weak convergence) in the set of all probability distributions on non-negative reals, and they can approximate any distribution with nonzero density in $(0, \infty)$ to any desired accuracy \cite{Cinneide99}. Moreover, the matrix representation of PH distributions makes them suitable for theoretical analysis. The PH distribution approach enables us to include more information about the characteristics of a jump process in its model, yet it preserves the analytical tractability of the exponential distribution. Then, one can employ powerful tools and techniques developed for Markovian models to analyze non-Markovian processes.  The PH distribution has various applications in reliability  and queueing theory \cite{He14}. It has been also used in \cite{Hou06, Li15} for stability analysis of phase-type semi-Markov jump linear systems.

Stability property and optimal control of Markov jump linear systems (MJLSs) have been extensively studied in the literature during the past decades under different assumptions of full-state feedback, output-feedback, completely and partially observable modes, and several control design issues have been discussed  \cite{Costa13, Vargas16, Dolgov16}.  For semi-Markov jump linear systems (S-MJLSs), stability and stabilization problems have been studied. More recently, in \cite{Zhang16, Zhang17}, stability properties of S-MJLSs are studied and numerically testable criteria for stability and stabilizability are provided. However, optimal control problem for general S-MJLS has not been adequately studied, to the best of our knowledge.

The main contributions of this paper are as follows. First, we introduce Markovianization-like techniques for non-exponential holding-time distributions from the realm of reliability theory to the domain of control design. Such approximations translate S-MJLS into a specific class of MJLSPOM. While control design for a general MJLSPOM has been studied before, our second contribution is in strengthening optimality conditions for such systems. In particular, we provide necessary and sufficient conditions for optimal linear controller for a general MJLSPOM. For the specific class of MJLSPOM obtained from S-MJLS, we additionally establish existence of optimal linear controller, as well as its optimality within a general class of controllers satisfying standard smoothness conditions. Third, by establishing that the optimal control gains depend only on the probability density functions of holding-time distributions, and not on a specific Markov-like representation, we consider pseudo-Markov representations. Such representations give lower computational complexity for optimal gain computation in comparison to their Markovian counterparts. 
Collectively, these contributions provide a novel set of tools for control design, and also to trade-off computational burden with control performance, for continuous-time S-MJLS.

The rest of the paper is organized as follows. Section~\ref{PRE1} gives preliminary definitions, notations, and technical results, used throughout the paper. Section~\ref{PS1} contains problem formulation for optimal control of S-MJLS. The Markovianization process using the PH distribution is outlined in Section~\ref{PH_Mark}. Optimal control results for MJLSPOM, including those specific to the context of S-MJLS, are presented in Section~\ref{OPTIMAL_CONTROL}. Section~\ref{JUMP_MR} discusses model reduction, and a pseudo-Markovianization representation using matrix exponential distribution, to reduce computation cost for control design.  To further illustrate the ideas presented in the paper, a numerical case study is given in Section~\ref{NUM_SIM}. Finally, concluding remarks are summarized in Section~\ref{CONC}.

\section{Preliminaries and Notations}\label{PRE1}

For a continuous random variable $T$, the \emph{probability density function} (pdf), the \emph{cumulative distribution function} (cdf), and the \emph{complementary cumulative distribution function} (ccdf)  are respectively denoted by $f_T(t)$, $F_T(t)$, and $\bar{F}_T(t)=1-F_T(t)$. The \emph{hazard rate function} of $T$ is defined as $h_T(t)=f_T(t)/\bar{F}_T(t)$. For the exponential distribution, the hazard rate function is constant. Associated with a (semi-) Markov process over a discrete state space $\mathcal{V}=\{1, 2, \ldots, m\}$, there is a directed graph  $\mathcal{G}=(\mathcal{V}, \mathcal{E})$ having vertex set $\mathcal{V}$ and edge set  $\mathcal{E}$. There is a directed arc from vertex $i$ to vertex $j$, denoted by $(i,j)\in \mathcal{E}$, if and only if direct transition from state $i$ to state $j$ is possible. The \emph{in-neighborhood} of state $i$ is defined as ${N}_i^-=\{j\in \mathcal{V}\,|\,(j,i)\in \mathcal{E}, j\neq i\}$, whose elements are called \emph{in-neighbors} of state $i$. Similarly, the \emph{out-neighborhood} of state $i$ is defined as ${N}_i^+=\{j\in \mathcal{V}\,|\,(i,j)\in \mathcal{E}, j\neq i\}$, whose elements are called \emph{out-neighbors} of state $i$. The probability that a Markov process is in state $i$ at time $t$ is denoted by $\mu_i(t)$. The \emph{mode indicator} of a random process is denoted by $\delta_{i}(t)$, which is equal to $1$ when the process is in mode $i$ at time $t$, and is $0$ otherwise. Then, $\mu_i(t)=\mathbb{E}[\delta_{i}(t)]$, where $\mathbb{E}[\cdot]$ denotes the expectation operator. The \emph{transition rate matrix} of a continuous-time homogeneous Markov chain is denoted by $\bar\Pi=[\pi_{ij}]$, where $\pi_{ij}$ is the rate at which transitions occur from state $i$ to state $j$, and $\pi_{ii}=-\sum_{j\neq i}\pi_{ij}$. The off-diagonal elements of $\bar\Pi$ are finite, non-negative, and the sum of all elements in any row of $\bar\Pi$ is zero. A state $i$ with $\pi_{ii}=0$ is called \emph{absorbing}, because the exit rate is zero and no transition can be fired from it. A non-absorbing state is called \emph{transient}.
Consider a time-homogeneous Markov chain with $m$ transient states and one absorbing state, and let $T$ be the time to enter the absorbing state from the transient states; then, the random variable $T$ is said to be \emph{phase-type (PH) distributed}. A PH distribution is represented by a triple  $(\Pi, \upeta, \upalpha)$, where $\Pi, \upeta, \upalpha$ have probabilistic interpretations in terms of a Markov chain as follows: (i) $\Pi\in\mathbb{R}^{m\times m}$ is referred to as the \emph{sub-generator matrix}, which is an invertible matrix with non-negative off-diagonal elements,  negative elements on the main diagonal, and non-positive row sums; the $ij$-th element, $i\neq j$, of $\Pi$ is the transition rate from transient state $i$ to transient state $j$,  (ii) ${\upeta}\in\mathbb{R}^m$ is called the  \emph{exit rate vector} (or the \emph{closing vector}), and satisfies ${\upeta}=-\Pi\mathds{1}_m$, where $\mathds{1}_m$ is a column vector with all elements equal to $1$; the vector $\upeta$ is element-wise non-negative and its $i$-th component is the transition rate from transient state $i$ to the absorbing state of the underlying Markov chain, and (iii) $\upalpha\in\mathbb{R}^m$ is called the \emph{starting vector}, which has non-negative elements, and satisfies $\upalpha^\top\mathds{1}_m\leq 1$; the $i$-th component of $\upalpha$ is the probability of being in transient state $i$ at the initial time  \cite[\S 1.2]{He14}. Each transient state of the underlying Markov chain of a PH distribution is referred to as a \emph{phase}.

\begin{lemma}\cite[\S 5.1]{Neuts89_a} \label{CDF_PDF}
Let $T$ be a non-negative random variable with an $m$-phase PH distribution represented by triple $(\Pi, \upeta, \upalpha)$. Then,
\begin{enumerate}[(i)]
  \item the pdf of $T$ is given by $f_T(t)=\upalpha^\top \text{exp}(\Pi t){\upeta}$, $t\geq 0$,  with Laplace transform $\mathcal{L}[f_T(t)]=\upalpha^\top(sI_m-\Pi)^{-1}{\upeta}$, where $I_m$ is the $m\times m$ identity matrix;
  \item the cdf of $T$ is given by
$F_T(t)=\mathbb{P}[T\leq t]=\int_0^tf_T(\tau)d\tau=1-\upalpha^\top\text{exp}(\Pi t)\mathds{1}_m$, $t\geq 0$, where $\mathds{1}_m$ is a column vector with all elements equal to $1$. Then, the ccdf  (or survival function) of $T$ is
$\bar{F}_T(t)=\mathbb{P}[T> t]=1-F_T(t)=\upalpha^\top\text{exp}(\Pi t)\mathds{1}_m$, $t\geq 0$; and
\item the $n$-th moment of  $T$ is $\mathbb{E}[T^n]=\int_0^\infty t^nf_T(t)dt=(-1)^nn!\upalpha^\top\Pi^{-n}\mathds{1}_m$.
\end{enumerate}
\end{lemma}

An $m$-state time-homogeneous continuous-time \emph{semi-Markov process} $\{r(t)\}$ is  described by three components: (i) an initial probability vector $\mu(0)\in\mathbb{R}^m$, where $\mu_i(0)=\mathbb{P}[r(0)=i]=\mathbb{E}[\delta_{i}(0)]$, (ii) a discrete-time \emph{embedded  Markov chain} with one-step transition probability matrix $P=[p_{ij}]\in\mathbb{R}^{m\times m}$ (with no self-loop, i.e., $p_{ii}=0$), which determines the mode to which the process will go next, after leaving mode $i$, and (iii) the conditional distribution function $F_{ij}(t)=\mathbb{P}[T_{ij}\leq t]$, where $T_{ij}$ is the time spent in mode $i$ from the moment the process last entered that mode, given that the next mode to visit is mode $j$ \cite[\S 9.11]{Stewart2009}. The random variable $T_{ij}$ is called a \emph{conditional holding time} of mode $i$. Hence, a semi-Markov jump process is completely specified by $(\mu(0), [p_{ij}], [F_{ij}])$. Sample paths of a semi-Markov process are specified as $(r_0, t_0), (r_1, t_1), (r_2, t_2), \ldots$, where the pair $(r_k, t_k)$ indicates that the process jumps to mode $r_k$ at time $t_k$  and remains there over the period $[t_k, t_{k+1})$.

Let $T_i$ denote the time spent in mode $i$ before making a transition (the successor mode is unknown). Then, $T_i=\sum_jp_{ij}T_{ij}$ with distribution function $\mathbb{P}[T_{i}\leq t]=\sum_j p_{ij}F_{ij}(t)$. The random variable $T_i$ is referred to as the \emph{unconditional holding time} of mode $i$. Obviously, if mode $j$ is the only out-neighbor of mode $i$, then $p_{ij}=1$ and $T_i=T_{ij}$.  In a semi-Markov process, once the system enters mode $i$, the process randomly selects the next mode $j\neq i$ according to the probability transition matrix $P=[p_{ij}]$. If mode $j$ is selected, the time spent in mode $i$ before jumping to mode $j$ is determined by the distribution function $F_{ij}(t)$.

For a function $f(t)$, $t\in[0,\infty)$, the $\ell_p$-norm, is defined as $\|f\|_p=\left(\int_0^\infty|f(t)|^pdt\right)^{1/p}$, for $p\in[1,\infty)$, and $\|f\|_p=\sup_t|f(t)|$, for $p=\infty$. Let $A=[a_{ij}]$ be an $m\times n$ matrix and $B=[b_{ij}]$ be a $p\times q$ matrix. The  \emph{Kronecker product} of $A$ and $B$ is an $mp\times nq$ matrix, defined as $A\otimes B=[(a_{ij}B)]$. The following lemma gives some properties of the Kronecker product.

\begin{lemma} \cite{Neudecker1969} \label{Kronecker_Prop}
The Kronecker product satisfies the following properties.
\begin{enumerate}[(i)]
  \item $(A\otimes B)(C\otimes D)=(AC)\otimes (BD)$, where  $A\in\mathbb{R}^{m\times n}$, $B\in\mathbb{R}^{p\times q}$, $C\in\mathbb{R}^{n\times s}$, and $D\in\mathbb{R}^{q\times r}$.
  \item  $\alpha(A\otimes B)=(\alpha A)\otimes B=A\otimes (\alpha B)$, where  $A\in\mathbb{R}^{m\times n}$, $B\in\mathbb{R}^{p\times q}$, and $\alpha$ is a scalar.
  \item $I_n\otimes A$ and $B\otimes I_m$ commute, for any $A\in\mathbb{R}^{m\times m}$ and $B\in\mathbb{R}^{n\times n}$.
  \item $\text{exp}(A\otimes I_s)=\text{exp}(A)\otimes I_s$, for any $A\in\mathbb{R}^{m\times m}$ and any positive integer $s$.
  \item Let $A, B$ be square symmetric matrices. If $A\succeq 0$ and $B\succeq 0$, then $A\otimes B\succeq 0$.
\end{enumerate}
\end{lemma}

For a linear state equation $\dot{X}(t)=A(t)X(t)$, $X(t_0)=X_0$, where $A(t)\in\mathbb{R}^{n\times n}$ is a bounded piecewise continuous function of  $t$, the unique continuously differentiable solution is $X(t)=\Phi(t, t_0)X_0$, where $\Phi_A(t, \tau)$ denotes the \emph{state transition matrix} associated with $A(t)$.

\begin{lemma}\label{STM1}
Let  the square matrices $M_1(t)$ and $M_2(t)$ be bounded piecewise continuous functions of $t$, with  state transition matrices $\Phi_{M_1}(t,\tau)$ and $\Phi_{M_2}(t,\tau)$, respectively.
\begin{enumerate}[(i)]
  \item For a block-diagonal matrix $M(t)=\text{diag}\left(M_1(t), M_2(t)\right)$,
the state transition matrix is given by $\Phi_M(t,\tau)=\text{diag}\left(\Phi_{M_1}(t,\tau), \Phi_{M_2}(t,\tau)\right)$. In general, if $M(t)=\text{diag}(M_1(t),$ $M_2(t),$  $\ldots,$ $M_n(t))$, then $\Phi_M(t,\tau)=\text{diag}(\Phi_{M_1}(t,\tau)$, $\Phi_{M_2}(t,\tau)$, $\ldots$, $\Phi_{M_n}(t,\tau))$.
  \item The state transition matrix of $M(t)=M_1(t)+M_2(t)$ is given by $\Phi_M(t,\tau)=$ $\Phi_{M_1}(t,0)$ $\Phi_{Z}(t,\tau)$$\Phi_{M_1}(0,\tau)$,  where
$Z(t)=\Phi_{M_1}(0,t)M_2(t)\Phi_{M_1}(t,0)$.
\end{enumerate}
\end{lemma}

\emph{Proof}: The proof is given in the Appendix.

\begin{lemma}\label{STM_SUM2}
Let $A\in\mathbb{R}^{m\times m}$ be a constant matrix and $B(t)\in\mathbb{R}^{n\times n}$ be a bounded piecewise continuous function of $t$. Then, the state transition matrix of $M(t)=(A\otimes I_n) + (I_m\otimes B(t))$ is given by
 $\Phi_M(t,\tau)=\text{exp}(A(t-\tau))\otimes \Phi_B(t, \tau)$.
\end{lemma}

\emph{Proof}: The proof is given in the Appendix.

\begin{lemma} \cite[\S 1.1]{Kandil03} \label{LEM3_sylvester}
Let $M(t)\in\mathbb{R}^{m\times m}$, $N(t)\in\mathbb{R}^{n\times n}$, and $U(t)\in\mathbb{R}^{m\times n}$ be bounded piecewise continuous functions of time $t$. The unique solution of the differential equation $\dot{X}(t)=M(t)X(t)+X(t)N(t)+U(t)$, $X(t_0)=X_0$, is given by
$X(t)=\Phi_M(t,t_0)X_0\,\Phi_{N^\top}^\top(t,t_0)+
\int_{t_0}^t\Phi_M(t,\tau)U(\tau)\Phi_{N^\top}^\top(t,\tau)d\tau$, $\forall t$,
where $\Phi_M(t,\tau)$ and $\Phi_{N^\top}(t,\tau)$ are the state transition matrices associated with square matrices $M(t)$ and $N^\top(t)$, respectively.
\end{lemma}

\begin{defn}\cite{Farina2000}\label{pos_sys_defn}
Consider a continuous-time LTI system with a rational transfer function.
\begin{enumerate}[(i)]
  \item Input-state-output positivity: Given a state-space representation of the system, if for any non-negative initial state and any non-negative input, the output and state trajectories are non-negative at all times, the system is said to be \emph{internally positive}.
  \item Input-output positivity:  Given the transfer function of the system, if the impulse response is non-negative at all times, the system is said to be \emph{externally positive}. In such systems, for any non-negative input, the output is always non-negative.
  \end{enumerate}
\end{defn}
Obviously, any internally positive system is also externally positive, but the converse is not true.

\begin{lemma}\cite{Farina2000}
An LTI system with a state-space realization $(A,b,c)$ is {internally positive} if and only if the off-diagonal elements of $A$ and all elements of $b, c$ are non-negative. A system that possesses such a realization is called \emph{positively realizable}.
\end{lemma}

\section{Problem Statement}\label{PS1}

Consider a continuous-time S-MJLS whose behavior over its utilization period $[0, t_{\text{f}}]$ is described by the following stochastic state-space model
\begin{equation}\label{plant}
\dot{x}(t)=A(r(t),t)x(t)+B(r(t),t)u(t),\;\;\;x(0)=x_0,
\end{equation}
where $t\in[0, t_{\text{f}}]$, the final time $t_{\text{f}}$ is finite, known and fixed, $x(t)\in\mathbb{R}^{{n}_x}$ is a measurable state vector,  $\{r(t)\}$ is a continuous-time semi-Markov process over a finite discrete state space that autonomously determines the mode of operation, and $u(t)\in\mathbb{R}^{{n}_u}$ is the control input. The signals $x(t)$ are $r(t)$ are respectively referred to as the continuous and discrete components of the system's state. If $r(t)=i$, we write $(A(r(t),t), B(r(t),t))=(A_i(t), B_i(t))$, where for each $i$, $A_i(t)$ and $B_i(t)$ are known, bounded, continuous and deterministic matrices representing the linearized model of the system at an operating point. The state of the jump process $r(t)$ is assumed to be observable, and statistically independent of $x(t)$, which are reasonable assumptions in many applications. For example, load level in power systems, wind speed, and solar irradiance can be measured online using sensing devices, and are independent of the continuous components of the system's state. Also, in modeling of an aircraft dynamics with multiple flight modes, when no information about the aircraft intent is available, the mode transitions are independent of the continuous dynamics \cite{Liu2011}. The optimal regulation problem is to find a control law of the form
\begin{equation}\label{control_law}
u(t)=\Gamma(r(t),t)x(t),
\end{equation}
where $\Gamma(r(t),t)$ is a gain matrix, such that, starting from a given initial condition $(x(0), r(0))$, the cost functional
\begin{equation}\label{cost}
\begin{split}
J=\mathbb{E}\Big{[}\int_{0}^{t_\text{f}}\!\!\left(x^\top\!(s)Q(r(s),s)x(s) + u^\top\!(s)R(r(s),s)u(s)\right)ds+ x^\top\!(t_\text{f})S(r(t_\text{f}))\,x(t_\text{f})\Big{]},
\end{split}
\end{equation}
subject to (\ref{plant}) and (\ref{control_law}), is minimized, where the weighting matrices $Q \succeq 0, S\succeq 0$, and $R\succ0$ can be mode-dependent. Linear feedback controllers, due to their simple structure and low complexity, are of practical interest; hence, it is desired to find the best controller in this class, in the sense that it optimizes a certain performance index. For simplicity, we assume that, in minimization of the cost functional (\ref{cost}), $x(t)$ and $u(t)$ are not constrained by any boundaries.

\begin{remark}
It is shown in Section~\ref{OPTIMAL_CONTROL} that the optimal control law  for problem (\ref{plant}), (\ref{cost}) (over all admissible control laws that satisfy some smoothness properties, and not just of the linear form in \eqref{control_law}) is in the form of switching linear state feedback (see Theorem~\ref{NS_conditions} and Remark~\ref{linearity_of_OPC}).
\end{remark}

\section{Markovianization of S-MJLSs Using the PH Distribution}\label{PH_Mark}

In order to deal with the control of S-MJLSs, a suitable model for the jump process is needed that accurately captures the characteristics of the actual process, yet retains the tractability of the control design problem. The PH distribution approach is a technique to exactly or approximately transform a semi-Markov process to a Markov process, to facilitate analysis of S-MJLSs. In order to Markovianize a semi-Markov process, the holding-time distribution of each mode is represented by a finite-phase PH model. For a mode with multiple out-neighbors, there are multiple conditional holding times with possibly different distributions. In this case, several PH models are to be designed, all corresponding to the same mode. This point is clarified by an example in the rest of this section.

\begin{remark}\label{PH_des}
In order to model a holding-time distribution, we consider PH models whose starting vector is of the form $\upalpha=[1, 0, \ldots, 0]^\top$. That is, the initial probability of the PH model is concentrated in the first phase. Such a model is often used in reliability theory when employing the Markov chain as a failure model for a component embedded in a larger system \cite{Cumani82}.
\end{remark}

\begin{figure}[h!]
\centering
\includegraphics[scale=0.3]{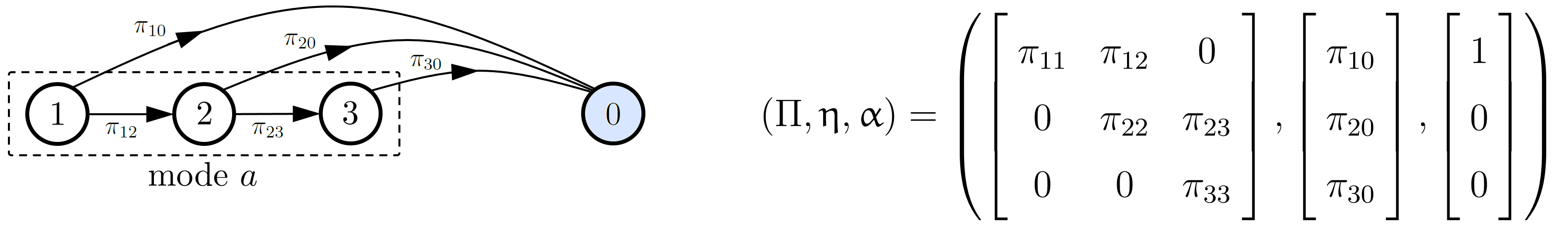}
\caption{A PH model (Cox) with three states $\{1, 2, 3\}$ that approximates the holding-time distribution of mode $a$ of a S-MJLS, where the arc labels represent the transition rates. The corresponding state-space realization is represented by the triple $(\Pi, \upeta, \upalpha)$.}
\label{fig_COX1}
\end{figure}

For simplicity of presentation, we consider a class of PH distributions called \emph{Coxian distribution}, whose sub-generator matrix has an upper bi-diagonal structure. The results, however, are applicable to any PH model as described in Remark~\ref{PH_des}. The Coxian distribution model, due to its simple structure and mathematical tractability, is often used in reliability theory for analysis and computation. Many PH distributions have a pdf-equivalent Coxian representation; for example, any PH distribution with triangular, symmetric, or tri-diagonal sub-generator matrix has a pdf-equivalent Coxian representation of the same order \cite[\S 1.4]{He14}. Moreover, the Coxian distribution is dense in the class of non-negative distributions  \cite{He14}. Figure~\ref{fig_COX1} shows the state transition diagram of a third-order Cox model, and the corresponding state-space representation.
From the parametric constraints of PH models, we have $\pi_{ii}<0$, $\pi_{ij}\geq 0$, $j\neq i$, $\pi_{11}=-(\pi_{12}+\pi_{10})$, $\pi_{22}=-(\pi_{23}+\pi_{20})$, $\pi_{33}=-\pi_{30}$. The pdf of the holding time of mode $a$ is given by $f_{a}(t)=\upalpha^\top\text{exp}(\Pi t)\upeta=\mathcal{L}^{-1}[\upalpha^\top(sI-\Pi)^{-1}{\upeta}]$, $t\geq 0$. Analogous to LTI systems where the transfer function (i.e., the Laplace transform of the impulse response) is unique while a state-space realization is not uniquely determined, any PH distribution has a unique pdf, but there is not a unique state-space representation $(\Pi, \upeta, \upalpha)$. Similarly, a PH model is called \emph{minimal}, if no pdf-equivalent PH model of smaller order exists. It can be easily verified that, in the model shown in Figure~\ref{fig_COX1}, if $\pi_{10}=\pi_{20}=\pi_{30}$, then the three-phase model is not minimal as it is pdf-equivalent to a single-phase model (i.e., exponential distribution) with pdf $f_a(t)=\pi_{30}\,\text{exp}(-\pi_{30}t)$, $t\geq 0$ .

\begin{figure}[h!]
\centering
\includegraphics[scale=0.43]{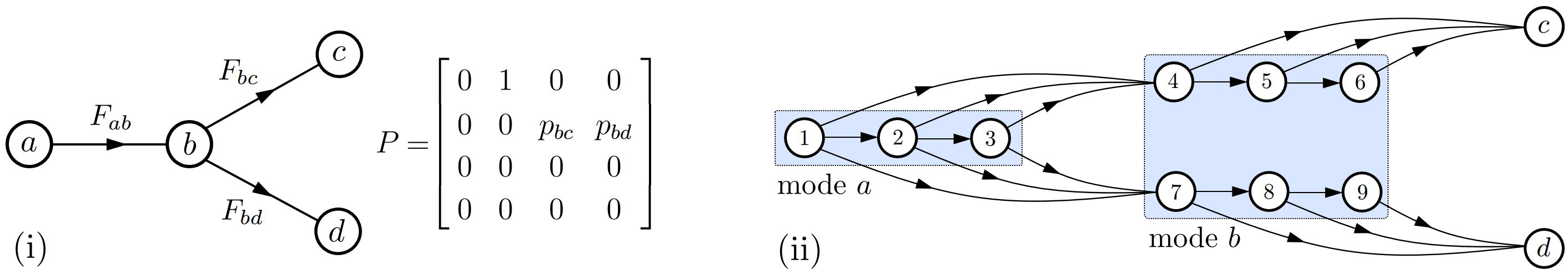}
\caption{(i) A four-mode semi-Markov process with general holding-time distributions and the one-step transition probability matrix of the corresponding embedded discrete-time Markov chain. (ii) A Markovianized version of the process, where each holding-time distribution is  approximated by a three-phase Cox model. In the Markovianized model, phases $1$-$3$ share the same dynamic as that of mode $a$ and phases $4$-$9$ share the same dynamic as that of mode $b$. The internal transitions of the Cox modes cannot be observed; only  transitions between modes $a, b, c, d$ are observable.}
\label{fig_COX3}
\end{figure}

In order to clarify PH-based Markovianization process, let us consider the four-mode semi-Markov process shown in Figure~\ref{fig_COX3}(i), where $F_{ab}$, $F_{bc}$, and $F_{bd}$ are conditional holding-time distributions, and $P$ is the one-step transition probability matrix of the corresponding embedded discrete-time Markov chain. The probabilities $p_{bc}$ and $p_{bd}$ in matrix $P$ can be computed as follows: $p_{bc}=\mathbb{P}[T_{bd}>T_{bc}]=\int_0^\infty \mathbb{P}[T_{bd}>t]f_{bc}(t)dt=\int_0^\infty (1-F_{bd}(t))f_{bc}(t)dt$ and $p_{bd}=1-p_{bd}$, where $f_{bc}(t)$ denotes the pdf of the holding time of mode $b$, given that the next mode to visit is mode $c$.
Since mode $b$ has two out-neighbors, if the process enters mode $b$, it jumps to mode $c$ after the time determined by $F_{bc}$, or jumps to mode $d$ after the time determined by $F_{bd}$. A Markovian approximation of the process by Coxian distributions is shown in Figure~\ref{fig_COX3}(ii), where each holding-time distribution is approximated by a three-phase model. The distributions $F_{ab}$, $F_{bc}$, and $F_{bd}$ are respectively approximated by states labeled $1$-$3$, $4$-$6$, and $7$-$9$. In each Cox model, all incoming links enter the first state of the model; however, the outgoing links may exit from any state of the model. If the process is initially in mode $b$  with probability $\mu_b(0)$, then in the Markovianized model, the process is initially in state $4$ with probability $\mu_4(0)= \mu_b(0)p_{bc}$ and in state $7$ with probability $\mu_7(0)=\mu_b(0)p_{bd}$, where $\mu_4(0)+\mu_7(0)=\mu_b(0)$. Let the exit rate vector of the Cox model of mode $a$ be denoted by $\upeta_a$. Then, in Figure~\ref{fig_COX3}(ii), the two vectors of transition rates from phases $\{1,2,3\}$ to phase $4$ and $7$ are respectively  $p_{bc}\upeta_a$ and $p_{bd}\upeta_a$.

When the underlying jump process of a S-MJLS is transformed to a Markov chain, all phases of each PH model associated with a particular mode share the same dynamic. For example, in the process shown in Figure~\ref{fig_COX3}, if $(A_a(t), B_a(t))$ and $(A_b(t), B_b(t))$ represent the dynamic of mode $a$ and $b$, respectively, then in Figure~\ref{fig_COX3}(ii), the dynamic of states $1$-$3$ is $(A_a(t), B_a(t))$, and that of states $4$-$9$ is $(A_b(t), B_b(t))$. It should be noted that, the transitions between the internal phases of a PH model cannot be observed or estimated; the sensing devices can only detect transitions between the modes of the semi-Markov process, i.e., only the jumps between modes $a, b, c, d$ can be observed. Therefore, a S-MJLS with completely observable modes is transformed to a specific MJLS with partially observable modes (MJLSPOM), i.e., where all the modes in a cluster have the same dynamic, the same weighting matrices, and the same control policy.

The PH distribution approach, however, suffers from a potential  drawback. Although in theory, any distribution on non-negative reals can be approximated arbitrarily well by a PH distribution, modeling of many distributions by PH models, with an acceptable level of accuracy, may need a very large number of phases. This can make design and analysis of S-MJLS computationally infeasible. For a wide class of distributions, the best PH approximate model of a reasonable size may result in an unacceptably large error in the distributions \cite{Fackrell05}. Hence, this approach may not allow us to accurately incorporate actual distribution functions into the jump process model. Therefore, it is necessary to find a compromise between modeling accuracy and the dimension of the model. We address this key limitation of the PH distribution approach in Section~\ref{JUMP_MR}, and propose a new technique for low-order modeling of non-exponential holding-time distributions.

\section{Optimal Control of S-MJLSs}\label{OPTIMAL_CONTROL}

Consider the problem formulated in Section~\ref{PS1}, and assume that the underlying semi-Markov jump process is replaced by a PH-based Markovianized model of an arbitrary large dimension. In this section, we present a control design procedure for a  MJLSPOM. Then, we investigate how the optimal controller and the cost value are related to the characteristics of the jump process.

\subsection{Optimal Control for MJLSPOM}
Consider a general MJLSPOM, i.e., where the modes in a cluster could have different dynamic, different weighting matrices, but not different control policy. 
It is assumed that, only transitions between the clusters can be observed, and no transition between the internal states of a cluster is observable. Then, associated with each cluster, a controller is to be designed, such that the cost functional (\ref{cost}) is minimized. The following theorem gives a necessary and sufficient condition for optimality of a linear state-feedback control law of the form (\ref{control_law}), for a general MJLSPOM.

\begin{theorem} \label{TM_Clustered_MJLS}
Consider a continuous-time MJLS of the form (\ref{plant}), and assume that the jump process $\{r(t)\}$ is a continuous-time homogeneous Markov chain with state space $\mathcal{V}=\{1, 2, \ldots, n_v\}$ and transition rate matrix $\bar\Pi=[\pi_{ij}]$. The system's dynamic in mode $i\in \mathcal{V}$ is represented by the pair $(A_i(t), B_i(t))$, and the transition rate from mode $i$ to mode $j$ is denoted by $\pi_{ij}$, $\forall i, j\in\mathcal{V}$. Assume that $\mathcal{V}$ is partitioned into $q$ disjoint subsets (clusters) $\mathcal{C}_1, \mathcal{C}_2, \ldots, \mathcal{C}_q\subseteq\mathcal{V}$, where $\bigcup_{i=1}^q\mathcal{C}_i=\mathcal{V}$, and that only transitions between clusters can be observed. Also, assume that the control law is of the form $\{u(t)=\Gamma_k(t)x(t)$, if $r(t)\in\mathcal{C}_k\}$. Suppose there exists a set of optimal gains $\{\Gamma_k(t), k=1,2,\ldots, q, t\in[0, t_{\text{f}}]\}$ that minimizes (\ref{cost}), for given $x(0)=x_0$, initial cluster $\mathcal{C}_{i_0}$ (i.e., $r(0)\in\mathcal{C}_{i_0}$), and initial probabilities $\mu_i(0)=\mathbb{P}[r(0)=i]$. Then, the optimal gains $\Gamma_k(t)$'s satisfy   (\ref{MJLS_opt_cond})-(\ref{MJLS_cov}):
\begin{equation}\label{MJLS_opt_cond}
\sum_{i\in \mathcal{C}_k}\left(R_i(t)\Gamma_k(t)+B_i^\top\!(t) \Lambda_i(t)\right)X_i(t)=0,
\end{equation}
for $k=1,2,\ldots,q$, and all $t\in[0, t_{\text{f}}]$, where $\Lambda_i(t)$ is the \emph{co-state matrix} of mode $i$ satisfying
\begin{equation}\label{MJLS_costate}
\begin{split}
-\dot{\Lambda}_i(t) &= \bar{A}_i^\top\!(t)\Lambda_i(t)+\Lambda_i(t)\bar{A}_i(t)+L_i(t)+\sum_{j\in \mathcal{V}}\pi_{ij}\Lambda_j(t),\;\;\;\Lambda_i(t_{\text{f}})=S_i,
\end{split}
\end{equation}
for all $i\in \mathcal{V}$, where  $\bar{A}_i(t)=A_i(t)+B_i(t)\Gamma_i(t)$ is the closed-loop matrix of mode $i$, $\Gamma_i(t)=\Gamma_k(t)$, $\forall i\in \mathcal{C}_k$, $L_i(t)=Q_i(t) + \Gamma_i^\top\!(t)R_i(t)\Gamma_i(t)$, and $X_i(t)=\mathbb{E}[x(t)x^\top\!(t)\delta_i(t)]$ is the \emph{covariance matrix} of mode $i$ which satisfies
\begin{equation}\label{MJLS_cov}
\dot{X}_i(t)=\bar{A}_i(t)X_i(t)+X_i(t)\bar{A}_i^\top\!(t)+\sum_{j\in \mathcal{V}}\pi_{ji}X_j(t),\;\;\;X_i(0)=x_0x_0^\top\mu_i(0),
\end{equation}
for all $i\in \mathcal{V}$, where $\delta_i(t)$ is the mode indicator function. Conversely, if (\ref{MJLS_opt_cond})-(\ref{MJLS_cov}) are satisfied, then $\Gamma_k(t)$'s are optimal gains. Moreover, for \emph{any} set of bounded piecewise continuous control gains $\{\Gamma_i(t), i\in\mathcal{V}, t\in[0,t_\text{f}]\}$,  the cost function (\ref{cost}) can be expressed as
\begin{equation}\label{nominalcost}
\begin{split}
J\!=\!\int_{0}^{t_\text{f}}\!\sum_{i\in \mathcal{V}}\text{tr}[L_i(s)X_i(s)]ds\!+\! \sum_{i\in \mathcal{V}}\text{tr}[S_{i}X_i(t_\text{f})]\!=\!\sum_{i\in \mathcal{V}} \text{tr}[\Lambda_i(0)X_i(0)]\!=\! x_0^\top \Big{(}\!\sum_{i\in \mathcal{V}}\!\mu_i(0)\Lambda_i(0)\!\Big{)}x_0.
\end{split}
\end{equation}
\end{theorem}

\emph{Proof}:  The proof is given in the Appendix.

\begin{remark} \label{admissible} \leavevmode 
\begin{itemize}
	\item[(i)] From (\ref{nominalcost}), to evaluate the cost for a given set of of control gains $\{\Gamma_i(t), i\in\mathcal{V}, t\in[0,t_\text{f}]\}$, we just need to solve the co-state equation (\ref{MJLS_costate}), numerically backward in time. However, to compute the optimal control gains, we have to solve a set of nonlinear coupled matrix differential equations (\ref{MJLS_opt_cond})-(\ref{MJLS_cov}). They can be solved using the iterative procedures proposed in the literature for this class of equations (see  \cite[\S3.6]{Mariton90}, \cite[\S 6.9]{Kandil03}, \cite{Vargas16}).
	\item[(ii)] In Theorem~\ref{TM_Clustered_MJLS}, the internal states of a cluster $\mathcal{C}_k$ may have different dynamics and weighting matrices, but they share the same control gain $\Gamma_k(t)$. It should be noted that, Theorem~\ref{TM_Clustered_MJLS} is a general result and includes, as its special cases, MJLSs with completely observable modes (if every cluster is a singleton, $\mathcal{C}_i=\{i\}$, $i\in\mathcal{V}$), and MJLSs with no observable modes (if there is a single cluster containing all modes, $\mathcal{C}_1=\mathcal{V}$).
	\item[(iii)] In the case that every transition in the Markov jump process is observable, the covariance matrices will not appear in the controller equation (\ref{MJLS_opt_cond}). This is  because, in this case, every cluster is a singleton $\mathcal{C}_i=\{i\}$; then, for (\ref{MJLS_opt_cond}) to hold for any  $X_i(t)$, the controller equation reduces to  $\Gamma_i(t)=-R_i^{-1}(t)B_i^\top(t) \Lambda_i(t)$. Using the stochastic dynamic programming approach, it has been proven \cite{Wonham70} that, for the all-mode observable case, the linear stochastic switching feedback law $u(t)=-R_i^{-1}(t)B_i^\top(t) \Lambda_i(t)x(t)$ is the optimal controller, not only over the class of linear state-feedback controllers, but also over all admissible control laws $\mathcal{U}$ that satisfy some smoothness conditions, namely $\mathcal{U}=\{u(t)\,|\,u(t)=\psi( x(t), r(t), t), |\psi(x, r, t)-\psi(\bar{x}, r, t)|\leq \kappa_0|x-\bar{x}|,\, \psi(x, r, t)\leq \kappa_1(1+|x|),$ $\forall x, \bar{x}, r, t,$ and some finite $\kappa_0$, $\kappa_1>0\}$.
	\item[(iv)] The problem of finite-horizon optimal control of discrete-time MJLSPOM (i.e.,  discrete-time dynamics with a discrete-time Markov chain) is studied in \cite{Vargas16}, and a necessary condition for optimality of a linear state-feedback control law is provided. In \cite{Jafari17}, by exhibiting a numerical example, it is shown that, in the discrete-time setting, the necessary optimality condition given in \cite{Vargas16} is not sufficient, in general.
\end{itemize}
\end{remark}

We now return to the original problem and use the above result for control of S-MJLSs. After PH-based Markovianization of a semi-Markov process, each cluster $\mathcal{C}_k$ in Theorem~\ref{TM_Clustered_MJLS} will correspond to a mode of the S-MJLS. Hence, a particular MJLSPOM is obtained, where all internal states of each cluster share the same dynamic, weighting matrices, and the same controller (but not the same co-state and covariance matrices). The following theorem gives a necessary and sufficient condition for optimality of a control law for this class of MJLSPOM.

\begin{theorem}\label{NS_conditions}
Consider the system described in Theorem~\ref{TM_Clustered_MJLS}. In addition, assume that all internal states of each cluster $\mathcal{C}_k$  share the same dynamic, i.e., $A_i(t)=A_k(t)$ and $B_i(t)=B_k(t)$, $\forall i\in\mathcal{C}_k$. Then, for any given $x(0)=x_0$, initial cluster $\mathcal{C}_{i_0}$ (i.e.,  $r(0)\in\mathcal{C}_{i_0}$), and initial probabilities   $\mu_i(0)=\mathbb{P}[r(0)=i]$, the optimal control law, in the sense that (\ref{cost}) is minimized, is in the form of a switching linear state feedback
\begin{equation}\label{MJLS_opt_cond_rev}
u(t)=\Gamma_k(t)x(t),\;\text{if}\; r(t)\in\mathcal{C}_k,\;\text{where}\;\Gamma_k(t)=-R_k^{-1}(t)B_k^\top\!(t)\sum_{j\in\mathcal{C}_k} \frac{\mu_j(t)}{\sum_{i\in\mathcal{C}_k}\mu_i(t)}  {\Lambda}_j(t),
\end{equation}
where $\Lambda_i(t)$ satisfies (\ref{MJLS_costate}), and  $\mu(t)=\mu(0)\,\text{exp}(\bar\Pi t)$ is the row probability vector of the jump process. Moreover, global existence of positive semi-definite matrices $\Lambda_i(t)$, $\forall t\in[0, t_{\text{f}}]$, that satisfy (\ref{MJLS_costate}), (\ref{MJLS_opt_cond_rev}) is guaranteed.
\end{theorem}

\emph{Proof}: The proof is given in the Appendix.

\begin{remark}\label{linearity_of_OPC}
The control law (\ref{MJLS_opt_cond_rev}) is optimal, not only over the class of linear state-feedback control laws, but also over all admissible control laws $\mathcal{U}$ defined in Remark~\ref{admissible}(iii). Moreover, the covariance matrix does not appear in (\ref{MJLS_opt_cond_rev}); hence to compute the optimal gains, we just need to numerically solve a set of coupled matrix Riccati equation, by integrating backward in time. These results are analogous to those of the all-mode observable case~\cite{Wonham70}.
\end{remark}

\subsection{Dependency of Control Performance on Holding-Time Distributions}

As mentioned earlier in Section~\ref{PH_Mark}, a PH distribution does not have a unique state-space realization, and, for a given semi-Markov process, there may exist many PH-based Markovianized models with different structure and parameters, which are equivalent. Hence, when a semi-Markov process is Markovianized, and is used for control design, it is desired to investigate how the behavior and properties of the closed-loop system may depend on the structure and parameters of the Markovianized model of the jump process. The question is whether the use of different realizations of the jump process model may affect the cost value and the optimal control signal.

\begin{defn}
Two PH-based Markovianized models of a semi-Markov process are said to be \emph{pdf-equivalent}, if their PH models corresponding to the same holding time have the same pdf.
\end{defn}

In the sequel, we show that replacing the Markovianized model of a jump process with any pdf-equivalent model does not change the cost value and optimal controllers. In other words, the cost value and optimal control gains are invariant with respect to the selection of the state-space realization of holding-time distributions, as long as the realizations correspond to the same pdf. We first show that, for a given set of control gains $\Gamma_i(t)$'s, $t\in[0, t_{\text f}]$, the cost value $J$ depends on the distribution models through their entire pdf, over the control horizon. Since the initial probability of any PH model is concentrated in the first phase, then from (\ref{nominalcost}) and (\ref{MJLS_costate}), it suffices to show that the co-state matrix corresponding to the first phase of each distribution model is invariant for any choice of pdf-equivalent Markovianized models. This fact is established in the following theorem. For simplicity of presentation, a two-mode S-MJLS is considered; the results, however, hold true for any S-MJLS.

\begin{theorem}\label{Tm11}
Consider a two-mode S-MJLS, as shown in Figure~\ref{fig5}(i). The dynamic, control gain, and weighting matrices associated with modes~$a$ and $b$ are, respectively, represented by $\{(A_1(t), B_1(t)), \Gamma_1(t), (Q_1(t), R_1(t), S_1)\}$ and $\{(A_2(t), B_2(t)), \Gamma_2(t), (Q_2(t), R_2(t), S_2)\}$. Suppose that the holding-time distributions of mode~$a$ and $b$ are represented by an $m$-phase PH model $(\Pi_a, \upeta_a, \upalpha_a)$ and a $p$-phase PH model $(\Pi_b, \upeta_b, \upalpha_b)$, respectively, as shown in Figure~\ref{fig5}(ii), where $m, p$ are arbitrary finite numbers.
\begin{figure}[h!]
\centering
\includegraphics[scale=0.30]{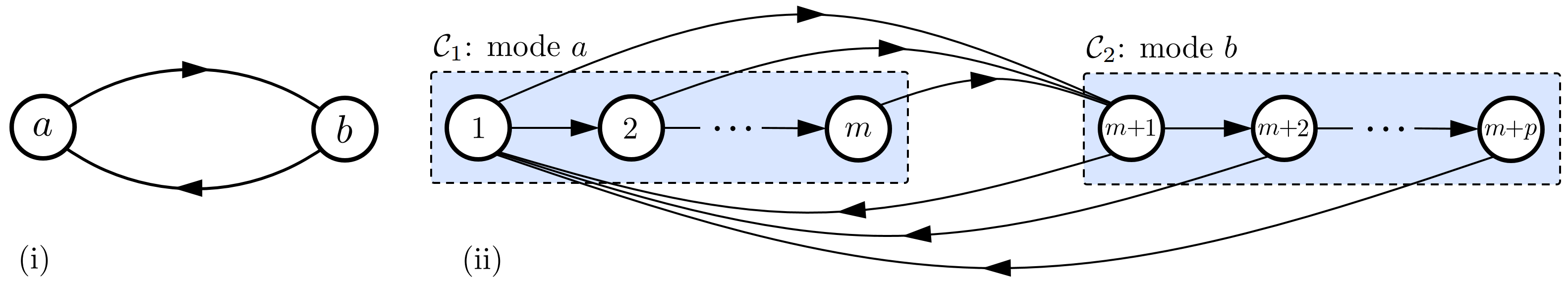}
\caption{(i) A two-mode S-MJLS, (ii) The holding-time distribution of each mode is represented by a PH model. All internal states of each model share the same dynamic, control gain, and weighting matrices.}
\label{fig5}
\end{figure}
The co-state matrix of the first state of the PH models satisfies
\begin{equation}\label{Lambda_1eq}
\begin{split}
\Lambda_1(t)
= \bar{F}_{a}(t_\text{f}-t)\Phi_{\bar{A}_1}^\top\!(t_\text{f}, t)S_1\Phi_{\bar{A}_1}\!(t_\text{f},t) &+
\int_t^{t_\text{f}}\bar{F}_{a}(\tau-t)\Phi_{\bar{A}_1}^\top\!(\tau,t)L_1(\tau)\Phi_{\bar{A}_1}\!(\tau,t)d\tau\\
&+
\int_t^{t_\text{f}}f_{a}(\tau-t)\Phi_{\bar{A}_1}^\top\!(\tau,t)\Lambda_{m+1}(\tau)\Phi_{\bar{A}_1}\!(\tau,t)d\tau,
\end{split}
\end{equation}
\begin{equation}\label{Lambda_1eq2}
\begin{split}
\Lambda_{m+1}(t)
= \bar{F}_{b}(t_\text{f}-t)\Phi_{\bar{A}_2}^\top\!(t_\text{f}, t)S_2\Phi_{\bar{A}_2}\!(t_\text{f},t) &+
\int_t^{t_\text{f}}\bar{F}_{b}(\tau-t)\Phi_{\bar{A}_2}^\top\!(\tau,t)L_2(\tau)\Phi_{\bar{A}_2}\!(\tau,t)d\tau\\
&+
\int_t^{t_\text{f}}f_{b}(\tau-t)\Phi_{\bar{A}_2}^\top\!(\tau,t)\Lambda_{1}(\tau)\Phi_{\bar{A}_2}\!(\tau,t)d\tau,
\end{split}
\end{equation}
where $f_{a}(t), f_b(t)$ are, respectively, the pdfs of holding times of modes $a$ and $b$, and $\bar{F}_{a}(t)$, $\bar{F}_{b}(t)$ are the corresponding ccdfs,   $\Phi_{\bar{A}_i}(t, \tau)$ is the state transition matrix associated with the closed-loop state matrix $\bar{A}_i(t)=A_i(t)+B_i(t)\Gamma_i(t)$, $L_i(t)=Q_i(t) + \Gamma_i^\top\!(t)R_i(t)\Gamma_i(t)$, and $\Gamma_i(t)$ is a given control gain.
\end{theorem}

\emph{Proof}: The proof is given in the Appendix.

\begin{remark} \leavevmode 
\begin{itemize}
\item[(i)] From (\ref{nominalcost}) and Theorem~\ref{Tm11}, it follows that the dependency of the cost value, for given control gains, on the holding-time distributions is through their entire pdf over the control horizon; hence, replacing the jump process model by any pdf-equivalent model keeps the cost value invariant. 
\item[(ii)] Another implication of Theorem~\ref{Tm11} is that, the use of a distribution model obtained by matching the first few moments may lead to a large error in the cost value. For example, the \emph{rate equivalent} (or \emph{insensitivity}) approach is a simple method of Markovianizing a semi-Markov process in which any holding-time distribution of a semi-Markov process is replaced by an exponential one of the same mean \cite{Singh80, Katoen01}. This approach is proposed to study the steady-state behavior of some class of semi-Markov processes; however, the resulting error in the transient behavior can be very large. The use of such approximations may cause large errors in the pdfs, and hence a drastic change in the cost value.
\item[(iii)] In Theorem~\ref{Tm11}, without loss of generality, a two-mode S-MJLS is considered.  By following the same steps as in the proof of Theorem~\ref{Tm11}, it is easy to verify that the above results are valid for any S-MJLSs whose holding-time distributions are modeled by finite-phase PH models. That is, in general, for any given control gains, the control cost depends on holding times through their pdf over the control horizon.
\end{itemize}
\end{remark}

Let us assume that mode $a$ is the initial mode of the S-MJLS shown in Figure~\ref{fig5}(i), and the actual pdf of the holding time of this mode is denoted by $f_a(t)$, which can be realized by a finite-order PH model. Suppose $\hat{f}_a(t)$ is an estimate of $f_a(t)$ represented by a low-order PH model. From  (\ref{nominalcost}) and (\ref{Lambda_1eq}), for given control gains, the error in the cost due to the error between $f_a(t)$ and $\hat{f}_a(t)$ is given by
\begin{equation}
\tilde J=J-\hat{J}=x_0^\top\tilde\Lambda_1(0)x_0,
\end{equation}
where $\tilde\Lambda_1(0)=\Lambda_1(0)-\hat{\Lambda}_1(0)$. Let $\tilde\Lambda_i(t)=\Lambda_i(t)-\hat{\Lambda}_i(t)$; then from Theorem~\ref{Tm11}, it follows that
\begin{equation}\label{ERROR_LAMBDA1}
\begin{split}
&\tilde\Lambda_1(t)
\!=\! -\left(\int_0^{t_\text{f}-t}\!\!\!\!\!\!\!\!\varepsilon_a(\sigma)d\sigma\!\!\right)\Phi_{\bar{A}_1}^\top\!(t_\text{f}, t)S_1\Phi_{\bar{A}_1}\!(t_\text{f},t)\!
-\!\!
\int_t^{t_\text{f}}\!\!\left(\int_0^{\tau-t}\!\!\!\!\!\!\!\!\varepsilon_a(\sigma)d\sigma\!\!\right)\Phi_{\bar{A}_1}^\top\!(\tau,t)L_1(\tau)\Phi_{\bar{A}_1}\!(\tau,t)d\tau\\
&\;\;\;\;\;+\!
\int_t^{t_\text{f}}\!\!\!\varepsilon_a(\tau-t)\Phi_{\bar{A}_1}^\top\!(\tau,t)\Lambda_{m+1}(\tau)\Phi_{\bar{A}_1}\!(\tau,t)d\tau\!\!+\!\!
\int_t^{t_\text{f}}\!\!\!\hat{f}_a(\tau-t)\Phi_{\bar{A}_1}^\top\!(\tau,t)\tilde\Lambda_{m+1}(\tau)\Phi_{\bar{A}_1}\!(\tau,t)d\tau,\\
&\tilde\Lambda_{m+1}(t)=\int_t^{t_\text{f}}f_{b}(\tau-t)\Phi_{\bar{A}_2}^\top\!(\tau,t)\tilde{\Lambda}_{1}(\tau)\Phi_{\bar{A}_2}\!(\tau,t)d\tau,
\end{split}
\end{equation}
where $\tilde\Lambda_i(t_{\text{f}})=0$, $\varepsilon_a(t)=f_a(t)-\hat{f}_a(t)$ is the pdf error, and $\int_0^t\varepsilon_a(\sigma)d\sigma=F_a(t)-\hat{F}_a(t)$ is the cdf error. It is obvious from (\ref{ERROR_LAMBDA1}) that, for a given approximate pdf for the holding time of mode $a$, the amount of change in the cost due to the error in the holding time pdf depends on the dynamic, control gains, and weighting matrices.

\begin{example}\label{Example_1}
Consider a two-mode S-MJLS, as shown in Figure~\ref{fig5}(i), with scalar dynamic. Let the holding time of mode $a$ before jumping to mode $b$ be denoted by $T_{a}$. Suppose $T_{a}$ has a non-exponential distribution represented by a $3$-phase PH model with an upper bi-diagonal sub-generator matrix $\Pi_a=[\pi_{ij}]$ with diagonal and supper-diagonal elements $\pi_{11}=-10$, $\pi_{22}=-5$, $\pi_{33}=-0.01$, $\pi_{12}=1$, and $\pi_{23}=1$. For simplicity, let us assume that the holding time of mode $b$ is exponentially distributed, with a rate parameter equal to $0.1$. The dynamic and weighting matrices of modes $a$ and $b$ are respectively $(A_1, B_1, Q_1, R_1, S_1)=(1,0.1,1,1,0)$ and $(A_2, B_2, Q_2, R_2, S_2)=(-10,10,1,1,0)$. The system is initially in mode $a$, the initial condition of the system is $x_0=1$, and constant control gains $\Gamma_1=-12$ and $\Gamma_2=-6$ are given for mode $a$ and $b$, respectively. Let $J$ be the cost corresponding to the actual semi-Markov process, in which $T_{a}$ has a $3$-phase PH distribution with pdf $f_{a}(t)$ and mean $\mathbb{E}[T_{a}]=-\upalpha_a^\top\Pi_a^{-1}\mathds{1}_3=2.12$. The cost value, computed by solving (\ref{MJLS_costate}) for the given control gains, is equal to $J=x_0^2\Lambda_1(0)$, where $\Lambda_1(t)$ is the co-state variable of the first state of the PH model of $T_{a}$.  In order to evaluate the effect of modeling error in the distribution of $T_{a}$ on the cost value, let $\hat{J}$ be the cost value for the case when $f_{a}(t)$ is replaced by an exponential pdf $\hat{f}_{a}(t)$ (i.e., a single-phase PH model) with the same statistical mean as that of $T_{a}$, i.e.,  $\hat{f}_{a}(t)=\lambda\,\text{exp}(-\lambda t)$ with $\lambda=1/\mathbb{E}[T_{a}]$.  For the given control gains and final time $t_{\text{f}}=30$ sec, we obtain  $J=23.08$ and $\hat{J}=166.55$. Hence, even though the first moment of the two distributions are exactly the same, the large error in modeling of the entire pdf of $T_{a}$ over the control horizon leads to about $620\%$ relative change in the control cost. Therefore, in general, performance evaluation  of a given controller on a nominal system with a low-order approximate jump process model may be highly erroneous. $\hfill\square$
\end{example}

\begin{theorem}\label{THEOREM_OPG}
The optimal control gains obtained by solving (\ref{MJLS_costate}), (\ref{MJLS_opt_cond_rev}) depend on the holding-time distribution models through their entire pdf over the control horizon, and hence are invariant for any choice of pdf-equivalent PH-based Markovianized models.
\end{theorem}

\emph{Proof}: The proof is given in the Appendix.


Theorem~\ref{THEOREM_OPG} implies that, if the PH model of each holding-time distribution is replaced by a different, yet pdf-equivalent PH model, the optimal gains remain unchanged. The presence of error in holding-time pdfs, however, may adversely affect control performance.

\begin{example}
Consider the system and parameters given in Example~\ref{Example_1}. For the actual system with pdf $f_{a}(t)$, the optimal cost value is $J^*=10.60$ which is obtained by solving (\ref{MJLS_costate}), (\ref{MJLS_opt_cond_rev}). Now, we replace $f_{a}(t)$ by an exponential pdf $\hat{f}_{a}(t)$ (of the same statistical mean). Then, the two-mode S-MJLS is approximated by a two-mode MJLS. We consider the approximated model as a nominal model, based on which optimal control gains are computed. Let the obtained optimal gains for the nominal model be denoted by ${\hat{\Gamma}}_1^*(t)$ and ${\hat{\Gamma}}_2^*(t)$, $t\in[0, t_{\text{f}}]$. If we apply the control law $\{u(t)={\hat{\Gamma}}_i^*(t)x(t)$, if $r(t)$ is in mode $i\}$, to the actual S-MJLS, the achieved cost is $\hat{J}=28.32$. That is, computing the gains based on the approximate model for the holding time of mode $1$ leads to about $167\%$ relative increase in the cost value. This performance degradation is due to the error between $f_{a}(t)$ and $\hat{f}_{a}(t)$ over the control horizon.  $\hfill\square$
\end{example}

\section{Jump Process Modeling and Model Reduction}\label{JUMP_MR}

As pointed out at the end of Section~\ref{PH_Mark}, when using PH-based Markovianization to determine optimal control gains, we face two conflicting requirements. In order to make the control design computationally feasible, and yet achieve a satisfactory level of performance, a model of reasonable size for the jump process is needed. In this section, we study model order reduction of a semi-Markovian jump process. The problem is first investigated within the framework of PH distributions. Then, a more general class of distributions is introduced for modeling of the jump process.

\subsection{Modeling by PH-Distributions: PH-Based Markovianization}
The problem of fitting PH distributions to empirical data and modeling of a general distribution by PH models is a complex non-linear optimization problem \cite{Telek94}. There has been much research done on developing numerical algorithms to fit PH distributions to empirical data containing a large number of measurements \cite{Okamura16, Thummler05, Okamura11}. The method of \emph{maximum likelihood estimation} due to its desirable statistical properties has been widely used to estimate parameters of probability distributions, and \emph{expectation maximization} algorithms have been developed  to find the maximum-likelihood estimate of the parameters of a distribution \cite{Okamura11}. Analogous to modeling of dynamical systems by LTI models, after fitting a model to empirical data (modeling step), we need to develop a new model by appropriately reducing the order of the full-order model (model-reduction step), to be used for analysis and control design. We define the problem of model reduction for PH distributions as follows.

\begin{defn}[PH Model Reduction]\label{PHMRP}
Given an $m$-phase PH model $(\Pi, \upeta, \upalpha)$ with pdf $f(t)$, find an $\hat{m}$-phase PH model $(\hat\Pi, \hat\upeta, \hat\upalpha)$ with pdf $\hat{f}(t)$, where $\hat{m}<m$, such that the distance (with respect to some norm) between $\hat{f}(t)$ and $f(t)$ is made as small as possible.
\end{defn}

The connection between PH distributions and internally positive LTI systems (see Definition~\ref{pos_sys_defn}) has been discussed in \cite{Commault03}. One may use this connection to deal with the problem of PH model reduction. The characterizations of PH distributions and positive LTI systems are given next.

\begin{theorem}\label{PH_Characteristics} \cite{Cinneide90}
A continuous probability distribution on $[0, \infty)$ with a rational Laplace transform is of \emph{phase type} if and only if (i) it has a continuous pdf $f(t)$, such that $f(t)>0$ for all $t>0$ (and $f(0)\geq 0$), and (ii) $\mathcal{L}[f(t)]$ has a unique negative real pole of maximal real part (possibly with multiplicity greater than one).
\end{theorem}

\begin{theorem}\label{Positive_Characteristics} \cite{Farina96}
An LTI system with impulse response $h(t)$ has a \emph{positive realization} if and only if (i) $h(t)>0$ for all $t>0$ (and $h(0)\geq 0$), and (ii) $\mathcal{L}[h(t)]$ has a unique negative real pole of maximal real part (possibly with multiplicity greater than one).
\end{theorem}

From Theorems~\ref{PH_Characteristics}, \ref{Positive_Characteristics} and Lemma~\ref{CDF_PDF}, it follows  that the pdf and cdf of a PH distribution are respectively equivalent to the impulse response and the step response of a BIBO stable\footnote{An LTI system is bounded-input bounded-output (BIBO) stable if and only if its impulse response is absolutely integrable.} positive LTI system with state-space realization $(\Pi, \upeta, \upalpha)$. Hence, positivity-preserving model reduction techniques can be employed to deal with the problem in Definition~\ref{PHMRP}. Since we are interested in minimizing the distance between the pdfs, the $\ell_2$-norm of $\varepsilon(t)=f(t)-\hat{f}(t)$ can be used as a metric to measure the quality of a reduced-order model. From Parseval's relation, minimizing the $\ell_2$-norm in the time domain is equivalent to minimizing the $\mathcal{H}_2$-norm of the error in the frequency domain, because $\|\varepsilon\|_2^2=\int_0^\infty |\varepsilon(t)|^2dt = \|\mathcal{E}(s)\|_2^2=(1/2\pi)\int_{-\infty}^{+\infty}\left|\mathcal{E}(j\omega)\right|^2d\omega$, where $\mathcal{E}(s)=\mathcal{L}[f(t)-\hat{f}(t)]$. Minimizing the $\mathcal{H}_2$-norm of a transfer function, however,  is a non-convex problem and finding a global minimizer is a hard task \cite{Gugercin08}. One approach to handle the problem is to formulate it as a $\gamma$-suboptimal $\mathcal{H}_2$ model reduction defined as follows.

\begin{defn}[$\gamma$-suboptimal $\mathcal{H}_2$ PH Model Reduction]\label{SOPHMRP}
Consider an $m$-phase PH model $(\Pi,$ $\upeta,$ $\upalpha)$ with pdf $f(t)$. For a given $\gamma>0$, find (if it exists) an $\hat{m}$-phase PH model $(\hat\Pi, \hat\upeta, \hat\upalpha)$ with pdf $\hat{f}(t)$, where $\hat{m}<m$, such that $\|f-\hat{f}\|_2<\gamma$.
\end{defn}

In order to deal with the problem in Definition~\ref{SOPHMRP}, one may employ positivity-preserving $\gamma$-suboptimal $\mathcal{H}_2$ model reduction techniques developed for LTI systems. In \cite{Feng10}, an LMI-based algorithm is proposed which can used to find a reduced-order PH model. MATLAB toolbox YALMIP with solver SeDuMi can be used to solve the LMIs. In general, however, LMI-based algorithms, due to their computational complexity, are not applicable to high dimensional models. It is, therefore, desired to develop more efficient techniques for modeling of non-exponential holding-time distributions.

As indicated previously, due to the parametric constraints of PH distributions (i.e.,  the constraints on the elements of $(\Pi, \upeta, \upalpha)$ given in Section~\ref{PRE1}), accurate approximation by PH models may result in a very high-order model, especially when the density function has abrupt variations or has minima close to zero \cite{Telek94}. Moreover, there are many distributions with rational Laplace transform that are not phase-type. For example,  there is no finite-phase PH model that exactly represents  distributions with pdfs  $f(t)=e^{-t}(t-1)^2$ and $f(t)=0.4e^{-t}(1+t+\sin t)$, because the first one violates condition (i) and the second one violates condition (ii) of Theorem~\ref{PH_Characteristics}.

In modeling of holding-time distributions, the primary objective is to accurately capture the behavior of holding times, while maintaining tractability of control design. The question that arises is whether it is necessary for the distribution model $(\Pi, \upeta, \upalpha)$ to have a probabilistic interpretation in terms of a true Markov chain. Relaxing the sign constraints of the PH distribution leads to a larger class of distributions called \emph{matrix-exponential (ME) distributions} \cite{Fackrell05}, that provides more flexibility to reduce the order of the distribution models. Indeed, in Section~\ref{OPTIMAL_CONTROL}, we used the PH distribution approach as a mathematical tool to model a jump process for the purpose of computing the optimal control gains and evaluating control performance. To accomplish these objectives, it is not necessary to force the transition rates of the distribution models to be non-negative. Hence, instead of the PH distribution, a more general and more flexible class of distributions can be employed to accurately model the jump process with a smaller state-space dimension.

\subsection{Modeling by ME-Distributions: Pseudo-Markovianization}

The matrix-exponential (ME) distribution is a generalization of the PH distribution and has exactly the same matrix representation as that of the PH distribution given in Lemma~\ref{CDF_PDF}. However, the sign constraints on the elements of $(\Pi, \upeta, \upalpha)$ are removed \cite{Fackrell05}. A distribution on $[0,\infty)$ is said to be an ME distribution, if its density $f(t)$ has the form $f(t)=\upalpha^\top \text{exp}(\Pi t){\upeta}$, $t\geq 0$, where $\Pi$ is an invertible matrix, and $\upeta$, $\upalpha$ are column vectors of appropriate dimension. The ME distributions are only subject to the requirement that they must have a valid probability distribution function, namely the pdf must be non-negative, $f(t)\geq 0$, and must integrate to one, $\int_0^\infty f(t)dt=-\upalpha^\top\Pi^{-1}\upeta=1$. Hence, ME distributions can approximate more complicated distributions at a significantly lower order compared to the PH distribution \cite[\S 1.7]{He14}.

\begin{remark}
A \emph{pseudo-Markov chain} is a Markov-like chain with possibly negative transition rates \cite{Booth68}. Then, one could call the process of holding-time distribution modeling by ME distributions  \emph{pseudo-Markovianization} --- a technique for low-order approximation of non-exponential holding-time distributions.
\end{remark}

\begin{example}
Consider $f_a(t)=e^{-t}(t-1)^2$, $t\geq 0$ with $\mathcal{L}[f_a(t)]=(s^2+1)/(s+1)^3$ as the pdf of holding time of mode $a$ of a S-MJLS. Since $f_a(1)=0$, from Theorem~\ref{PH_Characteristics}(i), this distribution cannot be realized by a finite-order PH model. Indeed, an LTI system with transfer function $H(s)=(s^2+1)/(s+1)^3$ is externally positive, but not positively realizable. This distribution, however, can be exactly represented by a third-order ME distribution.  A realization of this distribution and the corresponding state transition diagram is shown in Figure~\ref{fig_ME1}. Although some transition rates are negative and the model has no probabilistic interpretation in terms of a true Markov chain, it perfectly describes the holding-time distribution of mode $a$. Hence, it is a suitable model for computing the optimal control gains and the cost value.
\begin{figure}[h!]
\centering
\includegraphics[scale=0.39]{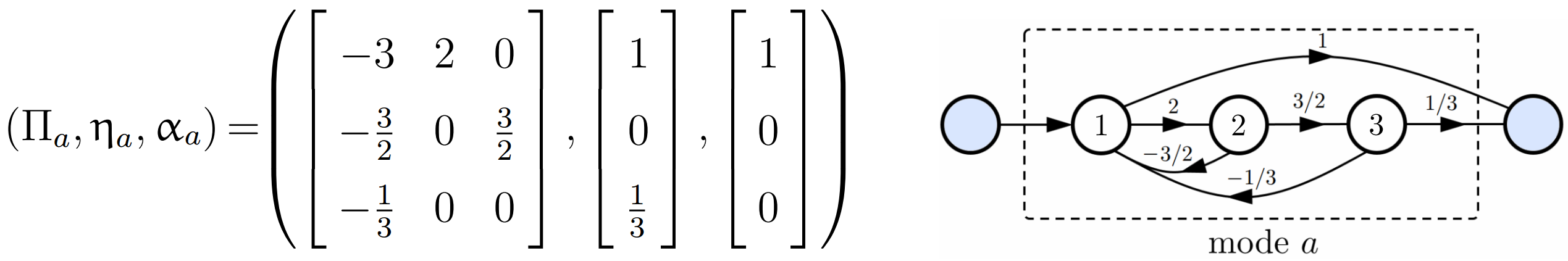}
\caption{Modeling of the holding-time distribution of mode $a$ with pdf $f_a(t)=e^{-t}(t-1)^2$, $t\geq 0$, by a $3$rd-order ME distribution. The corresponding state transition diagram does not represent a real Markov chain and is referred to as a pseudo-Markov chain.}
\label{fig_ME1}
\end{figure}
Fitting a third-order PH model using function `PHFromTrace'  \cite{Butools14} to a $50000$-sample data set obtained by inverse transform sampling gives a density function with about $30\%$ fit to the actual pdf $f(t)$, while the above third-order ME model gives a $100\%$ fit. The `fit percent' is defined in terms of the normalized root mean squared error expressed as a percentage, i.e.,  $\text{FitPercent}=100(1-\|f_a\!-\!\hat{f}_a\|/\|f_a\!-\!\bar{c}\|)$, where $f_a(t)$ and $\hat{f}_a(t)$ are time series of the actual pdf and the estimated pdf, respectively, the constant $\bar{c}$ is the arithmetic mean of $f_a$, and $\|\cdot\|$  indicates the Euclidean norm.$\hfill\square$
\end{example}

\begin{remark}
The results of Theorems~\ref{NS_conditions}, \ref{Tm11}, and \ref{THEOREM_OPG} are valid if PH models are replaced by ME models. Since, for any cluster $\mathcal{C}_k$, $\sum_{i\in\mathcal{C}_k}\mu_i(t)$ and $\sum_{i\in\mathcal{C}_k}\mu_i(t)\Lambda_i(t)$ depend on holding-time distribution models through their pdf and do not explicitly depend on transition rates (see the proof of Theorem~\ref{THEOREM_OPG}), then Theorem~\ref{NS_conditions} holds true for pseudo-Markov models. It is, also, shown in Theorems~\ref{Tm11} and \ref{THEOREM_OPG} that the control cost and optimal control gains depend on holding-time distribution models through their pdf. Hence, if each PH model is replaced with pdf-equivalent ME model, the cost value and optimal gains remain unchanged. It should be highlighted that, in a pseudo-Markov process, $\mu_i(t)$, for $t>0$, may be negative, and $\Lambda_i(t)$, for $t<t_{\text{f}}$, is not necessarily positive semi-definite, $\forall i$, however, for any cluster $\mathcal{C}_k$,  $\sum_{i\in\mathcal{C}_k}\mu_i(t)\geq 0$ $\sum_{i\in\mathcal{C}_k}\mu_i(t)\Lambda_i(t) \succeq 0$, and the co-state matrix associated with the first state of each ME model is positive semi-definite,  $\forall t\in[0, t_{\text{f}}]$.
\end{remark}

The fitting problem for the ME distribution is, however, very challenging \cite{Fackrell05}. The main difficulty is to ensure that the resulting ME representation has a non-negative density function. In the PH distribution, the sign constraints on $\Pi$ guarantee non-negativity of the density function; however, in the case of ME distribution, the sign constraints are relaxed and no simple criterion is available to determine whether a triple $(\Pi, \upeta, \upalpha)$ corresponds to a valid  distribution with a non-negative density. The problem of ME distribution fitting has been studied in several papers and a number of algorithms have been proposed. Moment matching methods are developed in \cite{Harris98, Liefvoort90}, however, they do not necessarily give a valid ME distribution. The function `MEFromMoments' in MATLAB toolbox \emph{Butools} \cite{Butools14} is based on the algorithm in \cite{Liefvoort90} which returns an ME distribution of order $N$ from a given set of $2N-1$ moments; the density function, however, is not guaranteed to be non-negative. A semi-infinite programming approach is proposed in \cite{Fackrell05, Fackrell12}, which requires some approximation in frequency domain to ensure that the result is a valid ME distribution; it is, however, not clear how the frequency domain approximation affects the time-domain behavior.

For modeling of each holding-time distribution in a semi-Markov process, we are looking for a realization $(\hat\Pi, \hat\upeta, \hat\upalpha)$ of the lowest possible order, such that  $\hat{f}(t)={\hat\upalpha}^\top\text{exp}(\hat\Pi t)\hat\upeta$ closely approximates the actual pdf of the holding time (and hence $\hat{F}(t)=1- {\hat\upalpha}^\top\text{exp}(\hat\Pi t)\mathds{1}$ closely approximates its cdf). For a distribution model with state-space representation $(\hat\Pi, \hat\upeta, \hat\upalpha)$, we make the following assumption:  (i) $\hat\Pi$ is Hurwitz, (ii) ${\hat\upeta}=-\hat\Pi\mathds{1}$, (iii) the starting vector is of the form $\hat\upalpha=[1, 0, \ldots, 0]^\top$, and (iv) $\hat{f}(t)={\hat\upalpha}^\top\text{exp}(\hat\Pi t)\hat\upeta\geq 0$, $\forall t\in[0, t_{\text{f}}]$. As is shown in the following lemma, assumptions (i)-(iii) are not restrictive constraints  for distribution modeling; the main difficulty is to ensure non-negativity of the density function.

\begin{lemma}\label{LTI_IMP}
Any BIBO stable LTI system with impulse response $h(t)$ and a strictly proper rational transfer function $H(s)=\mathcal{L}[h(t)]$ of minimal order $m$, with unit DC gain (i.e.,  $H(0)=\int_0^\infty h(t)dt=1$), can be represented by a triple $(\Pi, \upeta, \upalpha)$, where $\Pi\in \mathbb{R}^{m\times m}$ is a Hurwitz matrix,   ${\upeta}=-\Pi\mathds{1}_m\in \mathbb{R}^m$, and $\upalpha=[1, 0, \ldots, 0]^\top\in \mathbb{R}^m$. The impulse response of the system is expressed as $h(t)= \upalpha^\top\text{exp}(\Pi t)\upeta$, and the step response is $s(t)=\int_0^th(\tau)d\tau=1- \upalpha^\top\text{exp}(\Pi t)\mathds{1}_m$, $t\geq 0$.
\end{lemma}

\emph{Proof}: The proof is given in the Appendix.

Therefore, we need to find a rational transfer function $H(s)$ of minimal order $m$ which is: (i) strictly proper, (ii) stable, (iii) of unit DC gain, and (iv) externally positive, such that its impulse response $h(t)=\mathcal{L}^{-1}[H(s)]$ closely approximates the pdf of a given distribution. It should be noted that, $H(s)$ must be externally positive and does not need to be positively realizable. Even if $H(s)$ possesses a positive realization, we may not be interested in such a realization, because the dimension of a minimal positive realization may be much larger than $m$ which, in turn, unnecessarily increases the dimension of the jump process model. For example, the impulse response $h(t)=e^{-t}((t-1)^2+a)$, $t\geq 0$, is non-negative for any constant $a>0$, and the corresponding LTI model has a  third-order minimal rational transfer function which is positively realizable for any $a>0$; however, the order of its minimal positive realization goes to infinity as $a\rightarrow 0$ \cite{Cinneide90}.  It should be highlighted that the aforementioned four properties of $H(s)$ ensure that the step response $s(t)=\int_0^th(\tau)d\tau$ is a valid non-negative distribution function, i.e.,  a monotonically non-decreasing function starting at zero and approaching to one. In the sequel, we propose a procedure for fitting a ME distribution model to a class of life-time distributions.

Considering the connection between externally positive LTI systems and ME distributions, one can employ the sophisticated tools and techniques developed for LTI systems to deal with the problems of model fitting and model reduction for non-exponential distributions. For example, one may use available algorithms for transfer function fitting on time-domain input/output data. The pdf of a life-time  distribution can be viewed as the impulse response of an LTI system of finite or infinite order. Hence, a finite-order transfer function can be fitted to a sample time series data of the pdf. Some modifications, however, may be needed to make the impulse response of the model non-negative, over the control horizon. The following lemma gives sufficient conditions that guarantee the existence of a rational transfer function whose impulse response approximates a pdf with any desired accuracy.

\begin{lemma}\label{uniform_conv}
Consider a bounded, piecewise continuous, absolutely integrable function $f(t)$ defined in $[0, \infty)$. Let $H_n(s)$ denote an $n$th-order stable strictly proper rational transfer function and let $h_n(t)=\mathcal{L}^{-1}[H_n(s)]$. Then, there exists a sequence $\{h_n(t)\}$ that converges to $f(t)$ \emph{in the mean} as $n\rightarrow\infty$, that is $\int_0^\infty |f(t)-h_n(t)|^2dt\rightarrow 0$ as $n\rightarrow\infty$. If, in addition, $f(t)$ is differentiable and its time derivative is square integrable, then there exists a sequence $\{h_n(t)\}$ that \emph{uniformly} converges to $f(t)$ as $n\rightarrow\infty$, that is $\sup_t |f(t)-h_n(t)|\rightarrow 0$ as $n\rightarrow\infty$.
\end{lemma}

\emph{Proof}: The proof is given in the Appendix.

\begin{remark} \leavevmode 
\begin{itemize}
\item[(i)] The density functions of a wide class of life-time distributions satisfy the smoothness properties stated in Lemma~\ref{uniform_conv} (e.g. Erlang distribution, Weibull distribution with shape parameter $\kappa\geq 1$, Gamma distribution with shape parameter $\kappa\geq 1$, truncated normal distribution, etc.), hence they can be approximated uniformly by the impulse response of a BIBO stable LTI system with a strictly proper rational transfer function.
\item[(ii)] Since any density function $f(t)$ integrates to one, then from Lemma~\ref{uniform_conv}, it follows that $H_n(0)=\int_0^\infty h_n(t)dt\rightarrow 1$, as $n\rightarrow\infty$. For a finite $n$, the DC gain of the estimated transfer function can be set to one, by dividing the transfer function by its DC gain.
\item[(iii)] In order to fit a transfer function to a given density function, one may use MATLAB function `tfest(data, $n_p$, $n_z$)' from the System Identification Toolbox. This function fits a rational transfer function with $n_p$ poles and $n_z$ zeros to a given input/output time-domain data set. This function utilizes efficient algorithms for initializing the parameters of the model, and then updates the parameters using a nonlinear least-squares search method.
\end{itemize}
\end{remark}

It should be highlighted that Lemma~\ref{uniform_conv} does not guarantee non-negativity of $h_n(t)$. Even for large values of $n$, $h_n(t)$ may be slightly negative over some time intervals, or it may oscillate around zero. In the next subsection, two \emph{modifications} are proposed to make the obtained impulse response non-negative. Upper bounds on the resulting errors are also provided.

\subsection{Imposing the Non-negativity Constraint}

An impulse response $h_n(t)$ that provides a high fit percent to a pdf $f(t)$ may pass through the value of zero and violate the non-negativity constraint of density functions. Zero-crossings may occur over the time intervals where $f(t)$ is equal or very close to zero. We propose modifications so that the resulting impulse response fit is non-negative.

Let us split a given pdf $f(t)$ into two parts: (i) \emph{transient part} and (ii)  \emph{tail part}. The transient part is defined as $\{f(t)$, for $t\in[0, t_\epsilon]\}$, where $t_\epsilon$ is the time required for the ccdf, $\bar{F}(t)=1-\int_0^t f(\tau)d\tau$, to settle within a defined range $[0, \epsilon]$ near zero. Since $\bar{F}(t)$ is non-negative and monotonically decreasing, one could define $t_\epsilon$ as $t_\epsilon:=\{t\,|\,\bar{F}(t)=\epsilon\}$, where $\epsilon>0$ is some small constant. If one chooses $\epsilon=0.02$, then $t_\epsilon$ is referred to as the $2\%$ settling time of the distribution. The remaining part of $f(t)$ is called the tail part, i.e.,  $\{f(t), \text{for }t>t_\epsilon\}$. Zero-crossings in an approximation of a pdf may occur in either transient or tail part, or both. We propose two simple modifications to $f(t)$ in order to eliminate possible zero-crossings in the approximation $h_n(t)$.

In order to obtain a non-negative approximation to a pdf $f(t)$, we first give a procedure (Proposition~\ref{PROP1}) to obtain an impulse responses $h_n(t)$ that approximates $f(t)$, such that $h_n(t)$ has no zero-crossings for $t\in [0, t_\epsilon]$, $\forall n\geq n_0$, and some integer $n_0$. If the resulting approximation models have some zero-crossings for $t>t_\epsilon$, we need to apply another procedure (Proposition~\ref{PROP2}) to make the tail part non-negative, without introducing any zero-crossing in the transient part, and hence obtain a non-negative approximation to $f(t)$.

\begin{prop} \label{PROP1}
Consider a distribution with pdf $f(t)$ and settling time $t_\epsilon$, and assume that $f(t)$ satisfies the smoothness property given in Lemma~\ref{uniform_conv}. Let $h_n(t)$ denote the impulse response of an $n$th-order stable strictly proper rational transfer function $H_n(s)$ with unit DC gain. (i) If the transient part of $f(t)$ is bounded away from zero, i.e.,  $ f(t)\geq \gamma_0>0$, $\forall t\in[0, t_\epsilon]$, for some constant $\gamma_0$, then there exist an integer $n_0$ and a sequence $\{h_n(t)\}_{n\geq n_0}$ that closely approximates ${f}(t)$, $\forall t\geq 0$, such that $h_n(t)\geq 0$, $\forall t\in[0, t_\epsilon]$, $\forall n\geq n_0$. (ii) If the transient part of $f(t)$ is not bounded away from zero, there exists a smooth function $e(t)$, such that $\bar f(t)=(f(t)+e(t))/\|f+e\|_1$ is a valid pdf satisfying the smoothness property in Lemma~\ref{uniform_conv}, and is bounded away from zero for $t\in[0, t_\epsilon]$. Then, there exist an integer $n_0$ a sequence $\{h_n(t)\}_{n\geq n_0}$ that closely approximates $\bar{f}(t)$, $\forall t\geq 0$, such that $h_n(t)\geq 0$, $\forall t\in[0, t_\epsilon]$, $\forall n\geq n_0$. In addition, the distance (measured in $\ell_p$-norm, $p\in[1, \infty]$) between $h_n(t)$, $\bar{f}(t)$, and $f(t)$ satisfies
\begin{equation}\label{gtg456}
	\begin{split}
	\left|\|f-h_n\|_p-\|\bar f - h_n\|_p\right|\leq \frac{|1-\|f+e\|_1|}{\|f+e\|_1}\|f\|_p + \frac{1}{\|f+e\|_1}\|e\|_p.
	\end{split}
\end{equation}
\end{prop}

\emph{Proof}: The proof is straightforward and follows from the uniform convergence property given in Lemma~\ref{uniform_conv}; (\ref{gtg456}) follows from a simple application of the triangle inequality. The details of the proof are omitted for brevity. $\hfill\blacksquare$

\begin{remark} \leavevmode 
\begin{itemize}
\item[(i)] A simple example of $e(t)$, that locally pulls up $f(t)$ around $t=t_c$ is a bump function of the form $b(t)=a_0\,\text{exp}(-1/(d_0^2-(t-t_c)^2))$, for $t\in(t_c-d_0,t_c+d_0)$, and $b(t)=0$ elsewhere, for some $a_0, d_0>0$.
\item[(ii)] In Proposition~\ref{PROP1}(ii), a transfer function fitting algorithm tries to minimize the distance between $h_n(t)$ and $\bar f(t)$, while the true approximation error is between $h_n(t)$ and the actual pdf $f(t)$. Since for a large enough $n$,  a small-size $e(t)$ (in the $\ell_p$-norm sense) is needed to make $h_n(t)$ non-negative in $[0, t_\epsilon]$, then $h_n(t)$ will be also a good approximation of the actual pdf $f(t)$, $\forall t\geq 0$. This is because, any density function $f(t)$ satisfies $\|f\|_1=\int_0^\infty |f(t)|dt=1$. Then,  (\ref{gtg456}) implies that, as $\|e\|_p\rightarrow 0$, $f(t)+e(t)\rightarrow f(t)$, and hence $\|f+e\|_1\rightarrow 1$ and $\|f-h_n\|_p\rightarrow\|\bar{f} - h_n\|_p$.
\end{itemize}
\end{remark}

\begin{prop} \label{PROP2}
Consider a distribution with pdf $f(t)$ and settling time $t_\epsilon$, and assume that $f(t)$ satisfies the smoothness property given in Lemma~\ref{uniform_conv}. Let $h_n(t)$ denote the impulse response of an $n$th-order stable strictly proper rational transfer function $H_n(s)$ with unit DC gain. Let $\{h_n(t)\}_{n\geq n_0}$ be a sequence that approximates $f(t)$, such that $h_n(t)\geq 0$, $\forall t\in [0, t_\epsilon]$, $\forall n\geq n_0$, for some integer $n_0$. There exist positive reals $z_0, p_0$, and an integer $\bar{n}\geq n_0$, where $z_0\geq p_0>0$, such that  $\bar{h}_{n+1}(t):=\mathcal{L}^{-1}[W(s)H_n(s)]\geq 0$, $\forall t\geq 0$, $\forall n\geq \bar{n}$, where $W(s)=(p_0/z_0)(s+z_0)/(s+p_0)$. The approximation error between $\bar{h}_{n+1}(t)$ and $f(t)$ (measured in $\ell_p$-norm, $p\in[1, \infty]$) satisfies
\begin{equation}\label{gfg45676}
\begin{split}
\left|\|f-h_n\|_p-\|f - \bar{h}_{n+1}\|_p\right|\leq 2(1-p_0/z_0)\|h_n\|_p,
\end{split}
\end{equation}
where $\|h_n\|_p$ is finite for any $p\in[1, \infty]$.
\end{prop}

\emph{Proof}: The proof is given in the Appendix.

\begin{remark} \leavevmode 
\begin{itemize}
\item[(i)] Similar to compensation techniques in frequency domain, the selection of the best values for $p_0$ and $z_0$ is done by experience and trial-and-error. A general guideline is to place the pole of the compensator $W(s)$ at a reasonable distance to the right of the dominant pole of $H_n(s)$, such that $s=-p_0$ is the dominant pole of the compensated transfer function $W(s)H_n(s)$, and locate the zero of the $W(s)$ to the left of its pole.
\item[(ii)] The inequality (\ref{gfg45676}) implies that, as the distance between $z_0$ and $p_0$ goes to zero,  $\|f-\bar{h}_{n+1}\|_p\rightarrow\|f - h_n\|_p$. It should be highlighted that, for a large enough $n$, a small distance between $z_0$ and $p_0$ is needed to make $\bar{h}_{n+1}(t)$ non-negative. In this case,  $\bar{h}_{n+1}(t)$ is a good approximation of $f(t)$, $\forall t\geq 0$.
\end{itemize}
\end{remark}

In spite of the error introduced by the above modifications, numerical studies show that a modified model can approximate probability density functions more accurately compared to a PH model of the same order. Applying the above procedure to typical life-time distributions gives relatively low-order models with a good fit. This, in turn, enhances the control quality, without an unnecessary significant increase in complexity and computational burden. The following numerical example further illustrates the efficacy of the modifications proposed in Propositions~\ref{PROP1} and \ref{PROP2}.

\begin{example}\label{FITTING_EXAM}
Consider a Weibull random variable with pdf $f(t)=4t^3\text{exp}(-t^4)$, $t\geq 0$, and $2\%$ settling time $t_\epsilon=1.4$~sec. Let us first, fit a $5$th-order transfer function to a sample time series of $f(t)$ over time interval $[0, 100]$~sec, with sampling period of $10^{-3}$~sec. MATLAB function `tfest' gives a $5$th-order model whose impulse response has a $96.68\%$ fit to $f(t)$, as shown in Figure~\ref{fig_TF1}(i). However, both transient and tail parts of the obtained impulse response is slightly negative. To remove zero-crossings in the transient part, we apply Proposition~\ref{PROP1} as follows. Construct $\bar{f}(t)=f(t)+e(t)$ by slightly pulling up the initial part of $f(t)$, such that $\bar{f}(t)$ is bounded away from zero. Then, fit a $5$th-order transfer function $H_5(s)$ to $\bar{f}(t)$. Considering $e(t)=a_0\,\text{exp}(-1/(d_0^2-t^2))$, for $t\in[0, d_0)$, and $e(t)=0$, otherwise, with $a_0=30$ and $d_0=0.4$, we obtain a $5$th-order model whose impulse response $h_5(t)$ has a non-negative transient part; however, the tail part is oscillatory with many zero-crossings. In order to make $h_5(t)$ non-negative, we apply the filtering technique  in Proposition~\ref{PROP2}. Consider compensator $W(s)=k_0(s+0.85)/(s+0.8)$, where $k_0$ is chosen to make the DC gain of the compensated model $\bar{H}_6(s)=W(s)H_5(s)$ equal to one. Then, we have
$$\bar{H}_6(s)=\frac{0.1408s^5-2.037s^4+39.45s^3-274.6s^2+1629s+1609}{s^6+11.25s^5+109.7s^4+565.6s^3+1904s^2+3223s+1609},$$
whose impulse response $\bar{h}_6(t)$ is non-negative for all $t\geq 0$, and has a $92.70\%$ fit to $f(t)$. The performance can be improved by increasing the order of the model. To compare the quality of the compensated model with a PH model of the same order, we fit a $6$-phase PH distribution using MATLAB function `PHFromTrace' \cite{Butools14} to a $100000$-sample data set obtained by inverse transform sampling. The resulting PH model has a pdf with $60\%$ fit to $f(t)$. Figure~\ref{fig_TF1}(ii) shows $f(t)$, $\bar{h}_6(t)$, and the pdf of the $6$-phase PH distribution. $\hfill\square$
\begin{figure}[h!]
\centering
\includegraphics[scale=0.28]{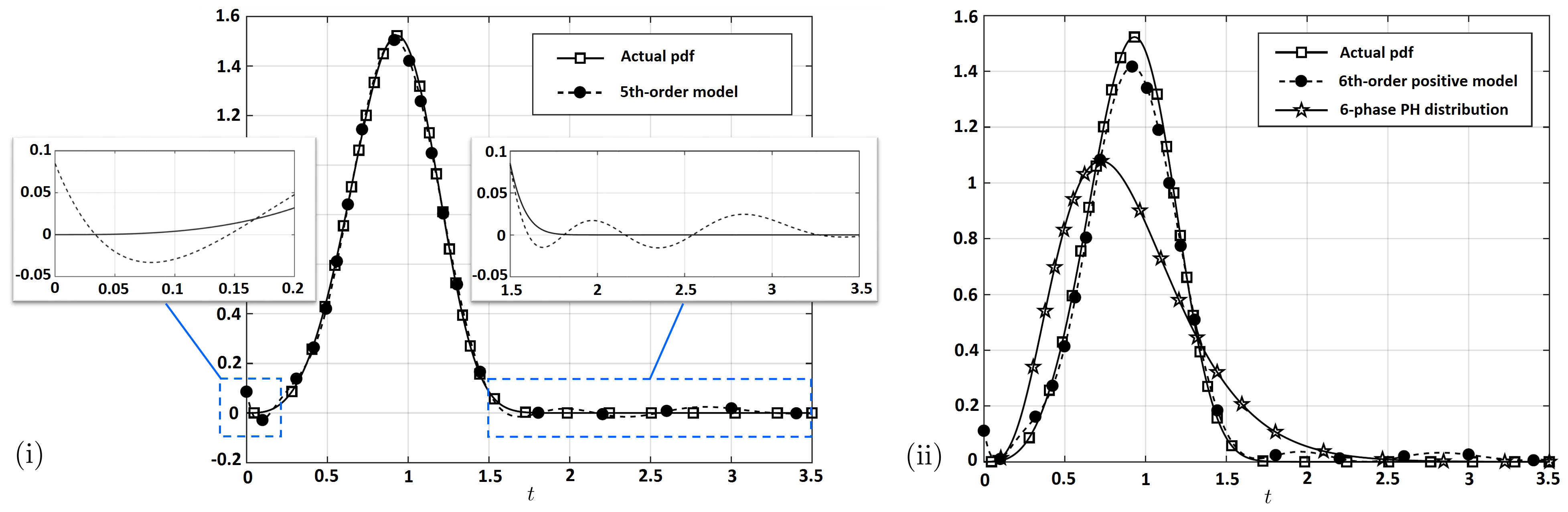}
\caption{Approximation of the pdf of a Weibull random variable. (i) The solid curve with square marker shows the actual pdf $f(t)$ and the dashed curve with circle marker is the impulse response of a $5$th-order LTI model (unconstrained) that has $96.68\%$ fit to $f(t)$, (ii) The curve with circle marker is the non-negative impulse of a $6$th-order LTI model with $92.70\%$ fit to $f(t)$. The curve with star marker is the pdf of a $6$-phase PH model with $60\%$ fit to $f(t)$.}
\label{fig_TF1}
\end{figure}
\end{example}

\section{Simulations}\label{NUM_SIM}

Power systems have nonlinear dynamics, and their operating conditions vary with the load level. A typical control design procedure is to partition the load range into several sub-ranges, each representing a mode of operation. A linear approximation model is then obtained associated with each mode \cite{Qiu04}. For a power system with randomly varying loads, a S-MJLS is well suited for describing the system's behavior.

\begin{figure}[h!]
\centering
\includegraphics[scale=0.26]{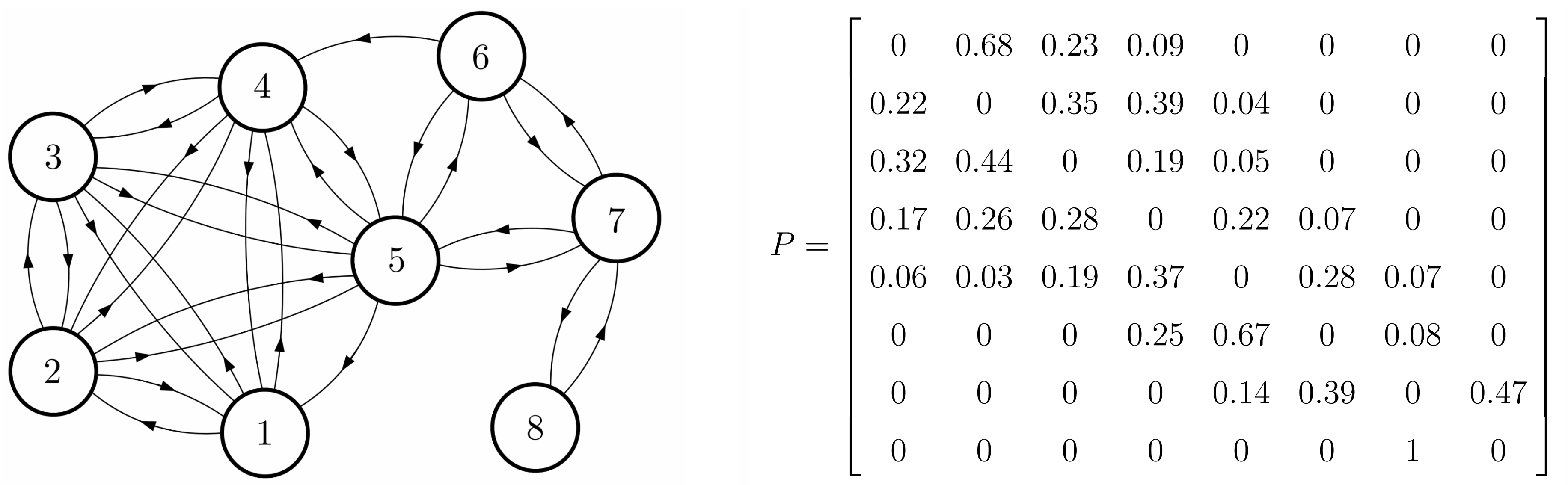}
\caption{State transition diagram of the semi-Markov model of a ship engine load process with eight operational modes, and the one-step transition probability matrix of the embedded Markov chain of the process.}
\label{fig_Engine2}
\end{figure}

Let us consider the load process of the ship engine in \cite[\S 8.2]{Grabski15}, which is modeled by a semi-Markov jump process. The load range is $[0, 3500]$ kW, which is partitioned into eight sub-ranges $[0, 250)$, $[250, 270)$, $[270, 280)$, $[280, 300)$, $[300, 350)$, $[350, 560)$, $[560, 1270)$, and $[1270, 3500]$ kW, each representing an operational mode of the engine. Figure~\ref{fig_Engine2} shows the state transition diagram of the load process, and the one-step transition probability matrix $P=[p_{ij}]$ of the embedded Markov chain of the semi-Markov process.
The elements of $P$ are obtained from empirical data as follows:  $p_{ij}=n_{ij}/\sum_kn_{ik}$, where $n_{ij}$ denotes the number of direct jumps from mode $i$ to mode $j$, $i\neq j$. Statistical analysis of the data indicates that  the Weibull distribution is a suitable model for the holding times of the process \cite{Grabski15}. For mode $i\in\{1, 2, \ldots, 8\}$, the cdf of the conditional holding times are $F_{ij}(t)=F_i(t)$, $\forall j\in N_i^+$, where $F_i(t)$ is the cdf of a Weibull distribution with shape parameter $\kappa_i$ and scale parameter $\lambda_i$, given in Table~\ref{TABLE1}. Using MATLAB function `tfest', we fit a transfer function of order $m_i$ to the holding-time pdf of each mode $i$, and evaluate the fitting quality. Some of the obtained models are modified to make the corresponding impulse responses non-negative. The normalized root mean squared errors between the actual and identified model for both pdfs and cdfs are listed in Table~\ref{TABLE1}.
\begin{table}[h!]
\centering
\caption{Fitting an externally positive, stable, strictly proper, rational transfer function of order $m_i$ and unit DC gain to Weibull distribution with parameters $\kappa_i$ and $\lambda_i$. The normalized root mean squared error between the actual and identified model measures the fitting quality.}
\label{TABLE1}
\begin{tabular}{ |c|c|c|c|c|c|c|c|c| }
 \hline
 mode & $1$ & $2$ & $3$ & $4$ & $5$ & $6$ & $7$ & $8$ \\\hline
 $\kappa_i$ (\text{shape}) & $2.5$ & $1.4$ & $1.2$ & $1.0$ & $1.8$ & $2.2$ & $2.0$ & $1.2$ \\
  $\lambda_i$ (\text{scale}) & $11.25$ & $8.99$ & $12.88$ & $11.34$ & $20.56$ & $9.01$ & $8.52$ & $15.54$ \\\hline\hline
 $m_i$ (\text{order}) & $4$ & $3$ & $3$ & $1$ & $3$ & $4$ & $4$ & $3$ \\\hline
 pdf Fit$\%$ & $98.35$ & $96.00$ & $98.00$ & $100$ & $98.79$ & $92.45$ & $96.49$ & $98.05$ \\\hline
 cdf Fit$\%$ & $99.64$ & $96.00$ & $99.03$ & $100$ & $99.19$ & $95.03$ & $96.89$ & $99.32$ \\\hline
\end{tabular}
\end{table}
Since for each mode, the conditional holding times are identically distributed, then we can transform the given eight-state semi-Markov process into a pseudo-Markov chain with $\sum_{i=1}^8 m_i=25$ states. For simplicity, we assign a scalar dynamic $(A_i, B_i)$ to each mode $i$ of the process, and compute the optimal control gains $\Gamma_i^*(t)$, $i\in\{1, 2, \ldots, 8\}$ and the corresponding cost $J(\Gamma^*)$. The system's parameters are $(A_1,B_1)=(0.5, -10)$, $(A_2,B_2)=(20, 0.2)$, $(A_3,B_3)=(10,1)$, $(A_4,B_4)=(5, -1)$, $(A_5,B_5)=(8, 2)$, $(A_6,B_6)=(4, 2)$, $(A_7,B_7)=(3, 1)$, $(A_8,B_8)=(5, -2)$, and weighting matrices are $Q_i=100I$, $R_i=100$, $S_i=0$, for $\forall i$, and $t_{\text{f}}=100$. We assume that the system is initially in mode $1$, and the initial state is $x_0=1$. From (\ref{MJLS_costate}), (\ref{nominalcost}), (\ref{MJLS_opt_cond_rev}),  we obtain $J(\Gamma^*)=11.63$. Now, let us assume that, in a nominal jump process model, each holding time is exponentially distributed with the same statistical mean as that of the corresponding Weibull distribution. Let the optimal control gains computed based on the nominal  model be denoted by $\hat{\Gamma}_i(t)$. By applying the control law $u(t)=\hat{\Gamma}_i(t)x(t)$ to the full-order pseudo-Markovianized model, the achieved cost is $J(\hat{\Gamma})=42.79$. That is, the modeling error of the jump process leads to $268\%$ relative increase in the cost. This demonstrates the importance of accurate modeling of the jump process for control of S-MJLSs.

\section{Conclusion}\label{CONC}

Optimal control of \emph{semi-}Markov jump linear systems is a relatively less studied topic in spite of several potential applications. This paper adopts a Markovianization approach to convert S-MJLS into MJLSPOM. 
%
%
While optimal control of general MJLSPOM has been studied previously, the fact that necessary conditions for optimal linear controller are also sufficient, as shown in this paper, appears to be novel, and hence could of independent interest. For MJLSPOM obtained from S-MJLS, an optimal linear controller is proven to exist, and is optimal within a general class of controllers. This is reminiscent of a similar result for MJLS whose all modes are observable. While phase-type approximation for holding times is commonly used in reliability theory, the use of matrix exponential approximation is relatively rare. This is potentially because the resulting pseudo-Markov representation does not have a meaningful probabilistic interpretation. However, the results in this paper suggest that such representations retain the required properties for control design, while lending computationally efficiency, and hence deserve further investigation.  

We plan to explore the proposed Markov-like approximations for other control settings that have traditionally been explored for MJLS. This includes output-feedback control in the presence of process and measurement noise, imperfect or delayed observation of the state of the jump process, and infinite-horizon control. Besides, the discrete-time setting poses new challenges. For example, in the companion paper~\cite{Jafari17}, we show that, unlike Theorem~\ref{TM_Clustered_MJLS} in this paper, the necessary condition is not sufficient, in general, in the discrete-time setting. 

\ifCLASSOPTIONcaptionsoff
\newpage
\fi
\bibliography{REFs}
\section*{Appendix}

\emph{Proof of Lemma~\ref{STM1}}: The proof follows by using the \emph{Peano-Baker series} \cite[\S 4]{Rugh96} or by showing that $\Phi_M(t,\tau)$ satisfies the equation ${\partial }\Phi_M(t,\tau)/{\partial t}=M(t) \Phi_M(t,\tau)$ with $\Phi_M(\tau,\tau)=I$, for any $t, \tau$. In the proof of (ii), the invertibility property of the state transition matrix is used, i.e.,  $\Phi_A^{-1}(t,\tau)=\Phi_A(\tau,t)$, for any $t, \tau$. $\hfill\blacksquare$

\emph{Proof of Lemma~\ref{STM_SUM2}}: From Lemma~\ref{STM1}(ii) with $M_1=A\otimes I_n$ and $M_2(t)=I_m\otimes B(t)$, we have, $\Phi_M(t,\tau)=\Phi_{M_1}(t,0)\Phi_{Z}(t,\tau)\Phi_{M_1}(0,\tau)$, where $Z(t)=\Phi_{M_1}(0,t)M_2(t)\Phi_{M_1}(t,0)$. Since $M_1$ is a constant matrix, then $\Phi_{M_1}(t,\tau)=\text{exp}(M_1(t-\tau))\overset{*}{=}\text{exp}(A(t-\tau)\otimes I_n)\overset{**}{=}\text{exp}(A(t-\tau))\otimes I_n$, where the identities ($*$) and ($**$) follow from Lemma~\ref{Kronecker_Prop}(ii) and Lemma~\ref{Kronecker_Prop}(iv), respectively. Then, we can write $Z(t) = \Phi_{M_1}(0,t)M_2(t)\Phi_{M_1}(t,0)= (\text{exp}(-At)\otimes I_n)(I_m\otimes B)(\text{exp}(At)\otimes I_n)\overset{\star}{=} (\text{exp}(-At)\otimes I_n)(\text{exp}(At)\otimes I_n)(I_m\otimes B)\overset{\star\star}{=} (I_m\otimes I_n)(I_m\otimes B)=I_m\otimes B$, where the identity ($\star$) follows from Lemma~\ref{Kronecker_Prop}(iii), and the identity ($\star\star$) follows from Lemma~\ref{Kronecker_Prop}(i) and the fact that for square matrices $A_1, A_2$, $\text{exp}(A_1)\text{exp}(A_2)=\text{exp}(A_1+A_2)$, if and only if $A_1$ and $A_2$ commute. Since $Z(t)=I_m\otimes B=\text{diag}(B, \ldots, B)$ is block diagonal,  from Lemma~\ref{STM1}(i), $\Phi_Z(t,\tau) = \text{diag}(\Phi_B(t,\tau), \ldots, \Phi_B(t,\tau)) = I_m\otimes \Phi_B(t,\tau)$. Then, $\Phi_M(t,\tau) = \Phi_{M_1}(t,0)\Phi_{Z}(t,\tau)\Phi_{M_1}(0,\tau) \!=\! (\text{exp}(At)\otimes I_n)(I_m\otimes \Phi_B(t, \tau))(\text{exp}(-A\tau)\otimes I_n) \!=\! (\text{exp}(At)\otimes I_n)(\text{exp}(-A\tau)\otimes I_n)(I_m\otimes \Phi_B(t, \tau)) \!=\! (\text{exp}(A(t-\tau))\otimes I_n)(I_m\otimes \Phi_B(t, \tau)) \!=\! \text{exp}(A(t-\tau))\otimes \Phi_B(t, \tau)$. $\hfill\blacksquare$

\emph{Proof of Theorem~\ref{TM_Clustered_MJLS}}: By taking differential of $X_i(t)=\mathbb{E}[x(t)x^\top\!(t)\delta_i(t)]$ and using the definition of mode indicator, it is easy to verify that $X_i(t)$ satisfies (\ref{MJLS_cov}), as $dX_i(t)=\mathbb{E}[d(x(t)x^\top\!(t))\delta_i(t)]+\mathbb{E}[x(t)x^\top\!(t)d\delta_i(t)]=(\bar{A}_i(t)X_i(t) + X_i(t)\bar{A}_i^\top\!(t))dt+\sum_{j\in\mathcal{V}}\pi_{ji}X_j(t)dt$ (see the proof of Theorem~3.5 in \cite{Mariton90}). We first show that (\ref{cost}) can be written as $J=\int_{0}^{t_\text{f}}\sum_{i\in \mathcal{V}}\text{tr}[L_i(s)X_i(s)]ds+ \sum_{i\in \mathcal{V}}\text{tr}[S_{i}X_i(t_\text{f})]$ as follows:
\begin{equation}\label{cost_new_form}
\begin{split}
J &= \mathbb{E}\Big{[}\int_{0}^{t_\text{f}}x^\top\!(s)L(r(s), s)x(s)ds+ x^\top\!(t_\text{f})S(r(t_\text{f}))\,x(t_\text{f})\Big{]}\\
&= \mathbb{E}\Big{[}\int_{0}^{t_\text{f}}\text{tr}[x^\top\!(s)L(r(s),s)x(s)]ds+ \text{tr}[x^\top\!(t_\text{f})S(r(t_\text{f}))\,x(t_\text{f})]\Big{]}\\
&= \mathbb{E}\Big{[}\int_{0}^{t_\text{f}}\text{tr}[L(r(s),s)x(s)x^\top\!(s)]ds+ \text{tr}[S(r(t_\text{f}))\,x(t_\text{f})x^\top\!(t_\text{f})]\Big{]}\\
&= \int_{0}^{t_\text{f}}\sum_{i\in \mathcal{V}}\text{tr}[L_i(s)X_i(s)]ds+ \sum_{i\in \mathcal{V}} \text{tr}[S_{i}X_i(t_\text{f})],
\end{split}
\end{equation}
where the second equality is because the cost functional is scalar and the trace of a scalar is itself, the third equality is obtained from the cyclic permutation invariance property of matrix trace, and the forth equality is due to the linearity of the expectation operator and that $L(r(s),s)x(s)x^\top\!(s)=\sum_{i\in\mathcal{V}}L_i(s)x(s)x^\top\!(s)\delta_i(s)$. Therefore, the stochastic optimization problem  (\ref{plant}), (\ref{cost}) is transformed into an equivalent deterministic one (\ref{MJLS_cov}), (\ref{cost_new_form}), in an average sense. 

\emph{Necessity}: The matrix minimum principle \cite{Athans67} can be applied to the deterministic optimization problem (\ref{MJLS_cov}), (\ref{cost_new_form}) to obtain a necessary optimality condition. The Hamiltonian function is given by $H=\sum_{i\in \mathcal{V}}(\text{tr}[L_iX_i]+ \text{tr}[\dot{X}_i\Lambda_i^\top])=\sum_{i\in \mathcal{V}}(\text{tr}[(Q_i+\Gamma_i^\top\! R_i \Gamma_i)X_i]+ \text{tr}[((A_i+B_i\Gamma_i)X_i + X_i(A_i+B_i\Gamma_i)^\top\!+\sum_{j\in\mathcal{V}}\pi_{ji}X_j)\Lambda_i^\top])$, where $\Lambda_i(t)$ is the co-state matrix associated with $X_i(t)$. For optimality of the control gains, the following conditions must hold for any $i\in \mathcal{V}$: (i) $-\dot{\Lambda}_i=\partial H/\partial X_i$, (ii) $\dot{X}_i=\partial H/\partial \Lambda_i$, and (iii) $\partial H/\partial \Gamma_k=0$, $k=1, 2, \ldots, q$. Using properties of trace and matrix derivatives \cite{Athans67}, conditions~(i) and (ii) lead to (\ref{MJLS_costate}) and (\ref{MJLS_cov}), respectively, and condition~(iii) yields (\ref{MJLS_opt_cond}). 

\emph{Sufficiency}: The dynamic programming approach \cite[\S 5]{Liberzon12} can be used to establish sufficiency. Let $X(t)=[X_1(t), \ldots, X_{n_v}(t)]^\top$, $L(t)=[L_1(t), \ldots, L_{n_v}(t)]^\top$, $S=[S_1, \ldots, S_{n_v}]^\top$, and  $\Gamma(t)=[\Gamma^\top_1(t), \ldots,$ $ \Gamma^\top_{q}(t)]^\top$. The cost functional can be expressed as $J(t_0, X_0)\!=\!\int_{t_0}^{t_\text{f}} \langle L(s), X(s)\rangle ds+ \langle S,X(t_{\text{f}})\rangle$, $X(t_0)=X_0$, where $t_0=0$, and $\langle E, F\rangle\!=\!\sum_{i} \text{tr}[E_i^\top\! F_i]$ is the inner product for the linear space of matrices $\{E=[E_1^\top, \ldots, E_{n_v}^\top]^\top, E_i\in\mathbb{R}^{n\times m}\}$.  Instead of minimizing $J(t_0, X_0)$ for given $t_0, X_0$, a family of minimization problems is considered with $J(t, X)=\int_{t}^{t_\text{f}} \langle L(s), X(s)\rangle ds+ \langle S, X(t_{\text{f}})\rangle$, $X(t)=X$, $t\in[t_0, t_{\text{f}})$. The optimal cost-to-go from $(t, X)$ is defined as $V(t,X)=\inf_{\Gamma[t, t_{\text{f}}]} J(t, X)$, $V(t_{\text{f}}, X)=\langle S, X\rangle$. Let $\dot{X}(t)=F(t, X(t), \Gamma(t))$, where $F_i(t, X, \Gamma)=(A_i+B_i\Gamma_i)X_i+X_i(A_i+B_i\Gamma_i)^\top+\sum_{j\in\mathcal{V}}\pi_{ji}X_j$. From the principle of optimality \cite[\S 5]{Liberzon12}, if a continuously differentiable function $V(t, X)$ (in both $X, t$) satisfies the Bellman's equation $\inf_{\Gamma\in\mathbb{R}^{qn_u\times n_x}}\left\{\langle L, X\rangle + \langle V_X(t,X), F(t,X, \Gamma) \rangle \right\}=-V_t(t,X)$, $V(t_{\text{f}}, X)=\langle S, X\rangle$, for all $t\in[0, t_{\text{f}})$ and all $X$, where $V_X(t,X)$ and $V_t(t,X)$ are the partial derivatives of $V$ with respect to $X$ and $t$, respectively, and if there exists an $qn_u\times n_x$ matrix $\Gamma$ minimizes the terms inside the brace, then $\Gamma$ is an optimal gain at $(t, X)$, and $V(t_0, X_0)$ is the optimal cost value for the process starting at $(t_0, X_0)$. Let us assume that the optimal cost-to-go is of the form $V(t, X)=\langle \Lambda(t), X\rangle = \sum_{i\in\mathcal{V}} \text{tr}[\Lambda_i(t)X_i]$, for some $\Lambda_i(t)\succeq 0$, $\forall t\in[0, t_{\text{f}}]$, with continuously differentiable elements, where $\Lambda(t)=[\Lambda_1(t), \ldots, \Lambda_{n_v}(t)]^\top$. For this function, the Bellman's equation can be expressed as $\inf_{\Gamma}\left\{\phi(\Gamma) \right\}=-\sum_{i\in\mathcal{V}}\text{tr}[\dot{\Lambda}_iX_i + Q_iX_i + \Lambda_iA_iX_i+\Lambda_iX_iA_i+\Lambda_i\sum_{j\in\mathcal{V}}\pi_{ji}X_j]$, where $\phi(\Gamma)=\sum_{i\in\mathcal{V}}\text{tr}[\Gamma_i^\top\!R_i\Gamma_iX_i + \Lambda_iB_i\Gamma_iX_i + \Lambda_iX_i\Gamma_i^\top B_i^\top]$.  We need to find a matrix  $\Gamma$ that minimizes $\phi$. Let the Jacobian of $\phi$, with respect to $\Gamma$, be denoted by $\mathbf{D}\phi$. Then, $\mathbf{D}\phi=2[\sum_{i\in\mathcal{C}_1}\left(R_i\Gamma_1X_i+B_i^\top\Lambda_iX_i\right),$ $\ldots, \sum_{i\in\mathcal{C}_q}\left(R_i\Gamma_qX_i+B_i^\top\Lambda_iX_i\right)]=0$, leads to (\ref{MJLS_opt_cond}). Moreover, from the definition of the Jacobian of a matrix function \cite{Magnus85}, we have  $\mathbf{D}^2\phi=2\,\text{diag}[\sum_{i\in\mathcal{C}_1}(X_i\otimes R_i), \ldots, \sum_{i\in\mathcal{C}_q}(X_i\otimes R_i)]$. Since $R_i\succ 0$ and $X_i\succeq 0$, then from Lemma~\ref{Kronecker_Prop}(v), $\mathbf{D}^2\phi\succeq 0$, $\forall \Gamma$, and hence $\phi$ is a convex function of $\Gamma$. Therefore, the obtained critical point is a global minimizer of $\phi$. It is easy to verify that, if $\Lambda_i(t)$'s satisfy (\ref{MJLS_costate}), then Bellman's equation is satisfied for any $X_i(t)$. Thus, a set of gains that satisfies  (\ref{MJLS_opt_cond})-(\ref{MJLS_cov})  is optimal, and the optimal cost is $V(0, X(0))= \sum_{i\in\mathcal{V}} \text{tr}[\Lambda_i(0)X_{i}(0)]$. Note that, for any set of bounded piecewise continuous control gains $\{\Gamma_i(t), t\in[0, t_{\text{f}}]\}$, equations   (\ref{MJLS_costate}) and (\ref{MJLS_cov}) have unique symmetric positive semi-definite solutions $X_i(t)$ and $\Lambda_i(t)$, $\forall t\in[0, t_{\text{f}}]$, respectively \cite[\S 3.3]{Costa13}, \cite{Wonham70}. Hence, $V(t, X)= \sum_{i\in\mathcal{V}} \text{tr}[\Lambda_i(t)X_i]$ is non-negative, for any  $t\in[0, t_{\text{f}}]$.  

In order to prove the last two identities in (\ref{nominalcost}), we post-multiply both sides of (\ref{MJLS_costate}) by $-X_i(t)$ and pre-multiply both sides of (\ref{MJLS_cov}) by $\Lambda_i(t)$. By adding them up, we obtain $L_iX_i=-d(\Lambda_iX_i)/dt + \Lambda_iX_i\bar{A}_i^\top\!-\!\bar{A}_i^\top\Lambda_iX_i + \sum_{j\in\mathcal{V}}(\pi_{ji}\Lambda_iX_j- \pi_{ij}\Lambda_jX_i)$. Since $X_i(t), \Lambda_i(t)$ are symmetric,   $\text{tr}[\Lambda_iX_i\bar{A}_i^\top]\!=\!\text{tr}[\bar{A}_i^\top\Lambda_iX_i]$, and $\sum_{i\in \mathcal{V}}\sum_{j\in\mathcal{V}}$ $(\pi_{ji}\Lambda_iX_j- \pi_{ij}\Lambda_jX_i)=0$, then $\sum_{i\in \mathcal{V}}\int_{0}^{t_\text{f}}\text{tr}[L_i(t)X_i(t)]dt
=\sum_{i\in \mathcal{V}}$ $(\text{tr}[\Lambda_i(0)X_i(0)]-\text{tr}[\Lambda_i(t_{\text{f}})X_i(t_{\text{f}})])$, where $\Lambda_i(t_{\text{f}})=S_i$. Therefore, $J=\sum_{i\in \mathcal{V}} \text{tr}[\Lambda_i(0)X_i(0)]$, and the last equality in (\ref{nominalcost}) follows from the cyclic permutation invariance property of matrix trace and that $X_i(0)=x_0x_0^\top\!\mu_i(0)$. $\hfill\blacksquare$

\emph{Proof of Theorem~\ref{NS_conditions}}: Consider the class of admissible  control laws $\mathcal{U}$ defined in Remark~\ref{admissible}(iii). The objective is to find $u(t)$, $t\in[t_0, t_{\text{f}}]$, such that for given $t_0$, $x(t_0)=x_0$, and initial cluster $\mathcal{C}_{k_0}$ (i.e.,  $r(t_0)\in \mathcal{C}_{k_0}$), the cost functional $J(t_0, x_0, u(\cdot), i_0)=\mathbb{E}[\int_{t_0}^{t_\text{f}}\mathcal{I}(s, x(s), u(s), r(s))ds+ x^\top\!(t_\text{f})S(r(t_\text{f}))\,x(t_\text{f})\,|\,x(t_0)\!=\!x_0, r(t_0)=i_0\in\mathcal{C}_{k_0}]$ is minimized, where $
\mathcal{I}(s, x(s), u(s), r(s))=x^\top\!(s)Q(r(s),s)x(s) + u^\top\!(s)R(r(s),s)u(s)$, and initial probability vector $\mu(t_0)=\mu_0$ is given. Using the stochastic dynamic programming approach \cite{Wonham70}, consider a family of minimization problems associated with the cost functional $J(t, x, u(\cdot), i)=\mathbb{E}[\int_{t}^{t_\text{f}}\mathcal{I}(s, x(s), u(s), r(s))ds+ x^\top(t_\text{f})S(r(t_\text{f}))\,x(t_\text{f})\,|\,x(t)=x, r(t)=i\in\mathcal{C}_k]$, where $t\in[t_0, t_{\text{f}})$ and $x(\cdot)$ in the integrand term  is a state trajectory satisfying $x(t)=x$ (a fixed value). It should be noted that, at time $t$, the probability vector is a given fixed value $\mu(t)=\mu$. The last argument $i$ of $J(t, x, u, i)$ denotes a mode in the initial cluster $\mathcal{C}_k$. It is not known which mode of $\mathcal{C}_k$ it is. It is only given that $i\in\mathcal{C}_k$, and that  the process at time $t$ is in mode $i$ with probability $\mu_i=\mathbb{P}[r(t)=i]$. Define the optimal cost-to-go from $(t, x, i\in\mathcal{C}_k)$ as $V(t,x, i)=\min_{u[t, t_{\text{f}}]} J(t, x, u, i)$, $V(t_{\text{f}}, x, i)=x^\top\!S_kx$. If a continuously differentiable scalar function $V$ (in both $t$ and $x$) is the solution to the following Bellman's equation, then it is the minimum cost for the process beginning at $(t,x,i\in\mathcal{C}_k)$, and the minimizing $u$ is the value of the optimal control at $(t,x,i\in\mathcal{C}_k)$:  $\min_{u\in\mathbb{R}^{n_u}}{\{}V_t(t,x, i)+ \mathfrak{L}_u V(t,x,i) + x^\top\!Q_kx+u^\top\!R_ku{\}}=0$, which must hold for any $t\in[t_0, t_{\text{f}})$, any $x$, and any probability vector $\mu$, where $\mathfrak{L}_u(\cdot)$ denotes the generator operator associated with the joint Markov process $\{(x(t),r(t))\}$ \cite{Wonham70}. When $\mathfrak{L}_u$ operates on $V(t,x,i)$, where $i\in\mathcal{C}_k$, it gives $\mathfrak{L}_u V(t,x,i)=V_x(t,x,i)^\top\! (A_kx+B_ku)+ \sum_{j\in\mathcal{V}} (\sum_{i\in\mathcal{C}_k} \mu_{i}\pi_{i j})V(t, x, j)$. It is obvious that, since $R_k\succ 0$, then term inside the brace in the Bellman's equation is a convex function of $u$, and hence attains it global minimum at $u=-(1/2)R_k^{-1}B^\top_kV_x(t, x, i)$. Let us assume that the optimal cost-to-go from $(t, x, i\in\mathcal{C}_k)$ is of the form $V(t, x, i)=\sum_{j\in\mathcal{C}_k} (\mu_j/\sum_{i\in\mathcal{C}_k}\mu_i) x^\top \Lambda_j(t)x$, for some $\Lambda_j(t)\succeq 0$, $\forall t\in[0, t_{\text{f}}]$, with continuously differentiable elements. Partial derivatives of $V(t, x, i)$ with respect to $t, x$ are $V_t(t, x, i)=\sum_{j\in\mathcal{C}_k} (\mu_j/\sum_{i\in\mathcal{C}_k}\mu_i) x^\top \dot{\Lambda}_j(t)x$ and $V_x(t, x, i)=\sum_{j\in\mathcal{C}_k} (2\mu_j/\sum_{i\in\mathcal{C}_k}\mu_i) {\Lambda}_j(t)x$. It is easy to verify that, if $\Lambda_i(t)$'s satisfy (\ref{MJLS_costate}), then the Bellman's equation is satisfied. 

In order to show the global existence of solution for the coupled Riccati equation (\ref{MJLS_costate}), (\ref{MJLS_opt_cond_rev}), we use the following two facts: (i) for any $i\in\mathcal{V}$, $\Lambda_i(t)$ is a symmetric positive semi-definite matrix, $\forall t\in[0, t_{\text{f}}]$, and (ii)   $V(t, x, i)=\sum_{j\in\mathcal{C}_k} (\mu_j/\sum_{i\in\mathcal{C}_k}\mu_i) x^\top \Lambda_j(t)x$ is the optimal cost-to-go from $(t, x, i\in\mathcal{C}_k)$, as long as it exists. Following the steps in \cite[\S 6.1.4]{Liberzon12}, it can be proved by contradiction that, no off-diagonal element of $\Lambda_i(t)$ exhibits a finite escape time (because otherwise, $\Lambda_i(t)$ is not positive semi-definite), and also no diagonal element of $\Lambda_i(t)$ can have a finite escape time (because otherwise, for some initial state $x$, the optimal cost $V(t,x,i)$ becomes unbounded, while the zero-input cost is finite). Therefore, the existence of the solution $\Lambda_i(t)$, $\forall i\in\mathcal{V}$, on the interval $[0, t_{\text{f}}]$ is guaranteed.   $\hfill\blacksquare$

\emph{Proof of Theorem~\ref{Tm11}}: Let $\Lambda_a=[\Lambda_1, \Lambda_2,$ $\ldots,$ $\Lambda_m]^\top$ and $\Lambda_b=[\Lambda_{m+1}, \Lambda_{m+2},$ $\ldots,$ $\Lambda_{m+p}]^\top$, where $\Lambda_i(t)\in\mathbb{R}^{n_x\times n_x}$. Then, (\ref{MJLS_costate}) can be expressed in terms of the Kronecker product as
\begin{subequations}\label{SSxc20}
\begin{align}
&\dot{\Lambda}_a = -(I_m\otimes\bar{A}_1^\top)\Lambda_a-(\Pi_a\otimes I_{n_x})\Lambda_a-\Lambda_a\bar{A}_1-(\upeta_a\otimes I_{n_x})\Lambda_{m+1}-(\mathds{1}_m\otimes L_1), \label{SSxc20_a}\\
&\dot{\Lambda}_b = -(I_p\otimes\bar{A}_2^\top)\Lambda_b-(\Pi_b\otimes I_{n_x})\Lambda_b-\Lambda_b\bar{A}_2-(\upeta_b\otimes I_{n_x})\Lambda_{1}-(\mathds{1}_p\otimes L_2), \label{SSxc20_b}
\end{align}
\end{subequations}
where $\Lambda_a(t_\text{f})=\mathds{1}_m\otimes S_1$ and $\Lambda_b(t_\text{f})=\mathds{1}_p\otimes S_2$. From (\ref{SSxc20}) and Lemma~\ref{LEM3_sylvester} we obtain
\begin{subequations}\label{Y_of_t}
\begin{align}
&\Lambda_a(t)=\Phi_{M_1}(t,t_\text{f})(\mathds{1}_m\otimes S_1)\,\Phi_{N_1^\top}^\top(t,t_\text{f})+
\int_{t_\text{f}}^t\Phi_{M_1}(t,\tau)U_1(\tau)\Phi_{N_1^\top}^\top(t,\tau)d\tau\\
&\Lambda_b(t)=\Phi_{M_2}(t,t_\text{f})(\mathds{1}_p\otimes S_2)\,\Phi_{N_2^\top}^\top(t,t_\text{f})+
\int_{t_\text{f}}^t\Phi_{M_2}(t,\tau)U_2(\tau)\Phi_{N_2^\top}^\top(t,\tau)d\tau
\end{align}
\end{subequations}
where $M_1(t)=-((\Pi_a\otimes I_{n_x})+(I_m\otimes\bar{A}_1^\top\!(t)))$, $N_1(t)=-\bar{A}_1(t)$, $U_1(t)=-((\upeta_a\otimes I_{n_x})\Lambda_{m+1}(t)+(\mathds{1}_m\otimes L_1(t)))$, $M_2(t)=-((\Pi_b\otimes I_{n_x})+(I_p\otimes\bar{A}_2^\top\!(t)))$, $N_2(t)=-\bar{A}_2(t)$, and  $U_2(t)=-((\upeta_b\otimes I_{n_x})\Lambda_{1}(t)+(\mathds{1}_p\otimes L_2(t)))$. Using the properties of state transition matrices  \cite[\S 1.1]{Kandil03}, we have $\Phi_{N_i^\top}^\top(t,\tau)=\Phi_{-\bar{A}_i^\top}^\top(t,\tau)=\Phi_{\bar{A}_i}(\tau,t)$, $\forall t, \tau$. From Lemma~\ref{STM_SUM2}, the state transition matrix of  $M_1(t)$ and $M_2(t)$ are respectively given by $\Phi_{M_1}(t,\tau)=
\text{exp}(\Pi_a(\tau-t))\otimes \Phi_{\bar{A}_1}^\top\!(\tau,t)$ and $\Phi_{M_2}(t,\tau)=
\text{exp}(\Pi_b(\tau-t))\otimes \Phi_{\bar{A}_2}^\top\!(\tau,t)$. Since $\Lambda_1(t)=(\upalpha_a^\top\otimes I_{n_x})\Lambda_a(t)$ and $\Lambda_{m+1}(t)=(\upalpha_b^\top\otimes I_{n_x})\Lambda_b(t)$, where $\upalpha_a=[1, 0, \ldots, 0]^\top\in\mathbb{R}^m$ and $\upalpha_b=[1, 0, \ldots, 0]^\top\in\mathbb{R}^p$, then from (\ref{Y_of_t}) we have
\begin{align}
&\Lambda_1(t) \!=\! (\upalpha_a^\top\!\!\otimes\! I_{n_x})\Phi_{M_1}\!(t,t_\text{f})(\mathds{1}_m\!\otimes \!S_1)\Phi_{\bar{A}_1}\!(t_\text{f},t)\!-\!\!
\int_t^{t_\text{f}}\!\!(\upalpha_a^\top\!\!\otimes\! I_{n_x})\Phi_{M_1}\!(t,\tau)U_1(\tau)\Phi_{\bar{A}_1}\!(\tau,t)d\tau, \label{Lambda_1eqpp_a}\\
&\Lambda_{m+1}(t) \!=\! (\upalpha_b^\top\!\!\otimes\! I_{n_x})\Phi_{M_2}\!(t,t_\text{f})(\mathds{1}_p\!\otimes \!S_2)\Phi_{\bar{A}_2}\!(t_\text{f},t)\!-\!\!
\int_t^{t_\text{f}}\!\!\!(\upalpha_b^\top\!\!\otimes\! I_{n_x})\Phi_{M_2}\!(t,\tau)U_2(\tau)\Phi_{\bar{A}_2}\!(\tau,t)d\tau.\label{Lambda_1eqpp_b}
\end{align}
From the expressions for $\Phi_{M_1}(t, t_\text{f})$ and $\Phi_{M_2}(t, t_\text{f})$, we have
\begin{equation*}
\begin{split}
(\upalpha_a^\top\otimes I_{n_x})&\Phi_{M_1}(t, t_\text{f})(\mathds{1}_m\otimes S_1) = (\upalpha_a^\top\otimes I_{n_x})(\text{exp}(\Pi_a(t_\text{f}-t))\otimes \Phi_{\bar{A}_1}^\top\!(t_\text{f},t))(\mathds{1}_m\otimes S_1) \\
&= (\upalpha_a^\top\text{exp}(\Pi_a(t_\text{f}-t))\mathds{1}_m)\otimes(\Phi_{\bar{A}_1}^\top\!(t_\text{f}, t)S_1)\overset{\ast}{=} \bar{F}_{a}(t_\text{f}-t)\Phi_{\bar{A}_1}^\top\!(t_\text{f}, t)S_1,
\end{split}
\end{equation*}
where the identity ($\ast$) follows from Lemma~\ref{CDF_PDF}. Similarly, $(\upalpha_b^\top\otimes I_{n_x})\Phi_{M_2}(t, t_\text{f})(\mathds{1}_p\otimes S_2) =  \bar{F}_{b}(t_\text{f}-t)\Phi_{\bar{A}_2}^\top\!(t_\text{f}, t)S_2$.
Using Lemma~\ref{Kronecker_Prop} we can write
\begin{equation*}
\begin{split}
&(\upalpha_a^\top\otimes I_{n_x})\Phi_{M_1}(t,\tau)U_1(\tau) = -(\upalpha_a^\top\otimes I_{n_x})(\text{exp}(\Pi_a(\tau-t))\otimes \Phi_{\bar{A}_1}^\top\!(\tau,t))U_1(\tau)\\
&= -(\upalpha_a^\top\text{exp}(\Pi_a(\tau-t))\otimes \Phi_{\bar{A}_1}^\top\!(\tau,t))((\upeta_a\otimes I_{n_x})\Lambda_{m+1}(\tau)+(\mathds{1}_m\otimes L_1(\tau)))\\
&= -((\upalpha_a^\top\text{exp}(\Pi_a(\tau-t))\upeta_a)\!\otimes\! \Phi_{\bar{A}_1}^\top\!(\tau,t) )\Lambda_{m+1}(\tau)\!-\!((\upalpha_a^\top\text{exp}(\Pi_a(\tau-t))\mathds{1}_m)\!\otimes\! (\Phi_{\bar{A}_1}^\top\!(\tau,t)L_1(\tau)))\\
&= -f_{a}(\tau-t)\Phi_{\bar{A}_1}^\top\!(\tau,t)\Lambda_{m+1}(\tau)-\bar{F}_{a}(\tau-t)\Phi_{\bar{A}_1}^\top\!(\tau,t)L_1(\tau),
\end{split}
\end{equation*}
and similarly $(\upalpha_b^\top\otimes I_{n_x})\Phi_{M_2}(t,\tau)U_2(\tau)=-f_{b}(\tau-t)\Phi_{\bar{A}_2}^\top\!(\tau,t)\Lambda_{1}(\tau)-\bar{F}_{b}(\tau-t)\Phi_{\bar{A}_2}^\top\!(\tau,t)L_2(\tau)$.
Substituting the above expressions in (\ref{Lambda_1eqpp_a}) and (\ref{Lambda_1eqpp_b}) leads to (\ref{Lambda_1eq}) and (\ref{Lambda_1eq2}). $\hfill\blacksquare$

\emph{Proof of Theorem~\ref{THEOREM_OPG}}: Without loss of generality, let us consider the system described in Theorem~\ref{Tm11}. From (\ref{MJLS_opt_cond_rev}), in order to prove the assertion of Theorem~\ref{THEOREM_OPG}, it suffices to show that for any cluster $\mathcal{C}_k$, $\sum_{i\in\mathcal{C}_k}\mu_i(t)$ and $\sum_{i\in\mathcal{C}_k}\mu_i(t)\Lambda_i(t)$ are invariant for any choice of pdf-equivalent Markovianized models. Let $\bar{\Pi}$ be the transition rate matrix of the overall PH-based Markovianized process. We have $\dot\mu(t)=\mu(t)\bar{\Pi}$, $\mu(0)=\mu_0$, where $\mu(t)=[\mu_a(t), \mu_b(t)]$, $\mu_a(t)=[\mu_1(t), \ldots, \mu_m(t)]$ and $\mu_b(t)=[\mu_{m+1}(t), \ldots, \mu_{m+p}(t)]$. The transient rate matrix $\bar{\Pi}$ can be written as
$$\bar{\Pi}=\left[\! \begin{array}{c:c}
\Pi_a & \upeta_a \upalpha_b^\top\\\cdashline{1-2}
\upeta_b \upalpha_a^\top & \Pi_b
\end{array}\! \right],$$   
 where $\upalpha_a=[1, 0, \ldots, 0]^\top\in\mathbb{R}^m$ and $\upalpha_b=[1, 0, \ldots, 0]^\top\in\mathbb{R}^p$. Then,
\begin{subequations}\label{mu_dot}
\begin{align}
&\dot{\mu}_a(t)=\mu_a(t)\Pi_a + w_b(t)\upalpha_a^\top,\;\;\mu_a(0)=\mu_1(0)\upalpha_a^\top,\\
&\dot{\mu}_b(t)=\mu_b(t)\Pi_b + w_a(t)\upalpha_b^\top,\;\;\mu_b(0)=\mu_{m+1}(0)\upalpha_b^\top,
\end{align} 
\end{subequations}
where $w_a(t)=\mu_a(t)\upeta_a$ and $w_b(t)=\mu_b(t)\upeta_b$. From (\ref{mu_dot}), we obtain
\begin{subequations}\label{mu_ab}
\begin{align}
&\mu_a(t) = \mu_1(0)\upalpha_a^\top\text{exp}(\Pi_a t)+\int_0^t w_b(\tau)\upalpha_a^\top \text{exp}(\Pi_a(t-\tau))d\tau, \label{mu_a}\\
&\mu_b(t) = \mu_{m+1}(0)\upalpha_b^\top\text{exp}(\Pi_b t)+\int_0^t w_a(\tau)\upalpha_b^\top \text{exp}(\Pi_b(t-\tau))d\tau. \label{mu_b}
\end{align} 
\end{subequations}
Post-multiplying both sides of (\ref{mu_a}) and (\ref{mu_b}) respectively by $\mathds{1}_m$ and $\mathds{1}_p$ gives
\begin{subequations}\label{RTE23}
\begin{align}
&\sum_{i\in\mathcal{C}_{1}}\mu_i(t)=\mu_a(t)\mathds{1}_m = \mu_1(0)\bar{F}_a(t)+\int_0^t w_b(\tau)\bar{F}_a(t-\tau)d\tau, \\
&\sum_{i\in\mathcal{C}_{2}}\mu_i(t)=\mu_b(t)\mathds{1}_p = \mu_{m+1}(0)\bar{F}_b(t)+\int_0^t w_a(\tau)\bar{F}_b(t-\tau)d\tau. 
\end{align} 
\end{subequations}
Similarly, post-multiplying both sides of (\ref{mu_a}) and (\ref{mu_b}) respectively by $\upeta_a$ and $\upeta_b$ gives
\begin{subequations}\label{RTE24}
\begin{align}
&w_a(t) = \mu_1(0)f_a(t)+\int_0^t w_b(\tau)f_a(t-\tau)d\tau, \\
&w_b(t) = \mu_{m+1}(0)f_b(t)+\int_0^t w_a(\tau)f_b(t-\tau)d\tau. 
\end{align} 
\end{subequations}
From (\ref{RTE23}) and (\ref{RTE24}), it follows that, $\sum_{i\in\mathcal{C}_k}\mu_i(t)$ is invariant for any choice of pdf-equivalent Markovianized models. 

Now, let us define $\xi_a(t)=\sum_{i\in \mathcal{C}_1} \mu_i(t)\Lambda_i(t)=(\mu_a(t)\otimes I_{n_x})\Lambda_a(t)$, where $\Lambda_a(t)$  satisfies (\ref{SSxc20_a}). Then, $\dot{\xi}_a(t)=(\dot\mu_a(t)\otimes I_{n_x})\Lambda_a(t) + (\mu_a(t)\otimes I_{n_x})\dot\Lambda_a(t)$.
From (\ref{SSxc20}), (\ref{mu_dot}), and Lemma~\ref{Kronecker_Prop}, we have
\begin{align}\label{xi_a}
&\dot{\xi}_a(t) = -\bar{A}_1^\top\!(t)\xi_a(t)-\xi_a(t)\bar{A}_1(t)+w_b(t)\Lambda_1(t)-w_a(t)\Lambda_{m+1}(t)-(\mu_a(t)\mathds{1}_m)L_1(t), 
\end{align} 
where $\xi_a(t_{\text{f}})=\sum_{i\in \mathcal{C}_1} \mu_i(t_{\text{f}})\Lambda_i(t_{\text{f}})=S_1\sum_{i\in \mathcal{C}_1} \mu_i(t_{\text{f}})=S_1\mu_a(t_{\text{f}})\mathds{1}_m$. The following identities are used to derive (\ref{xi_a}): $\mu_a(t)\otimes\bar{A}_1^\top\!(t)=\bar{A}_1^\top\!(t)(\mu_a(t)\otimes I_{n_x})$ and $\Lambda_1(t)=(\upalpha_a^\top\otimes I_{n_x})\Lambda_a(t)$. It is shown in Theorem~\ref{Tm11} that, for any control gains, $\Lambda_1(t)$ and $\Lambda_{m+1}(t)$ satisfy (\ref{Lambda_1eq}) and (\ref{Lambda_1eq2}). Moreover, it is shown in (\ref{RTE23}) and (\ref{RTE24}) that, $w_a(t)$, $w_b(t)$, and $\mu_a(t)\mathds{1}_m=\sum_{i\in\mathcal{C}_{1}}\mu_i(t)$ are invariant for any choice of pdf-equivalent Markovianized models; therefore, so is $\xi_a(t)$. Similarly, it can be shown that, $\xi_b(t)=\sum_{i\in \mathcal{C}_2} \mu_i(t)\Lambda_i(t)$ is invariant for any choice of pdf-equivalent Markovianized models. Hence, from (\ref{MJLS_opt_cond_rev}), the assertion of Theorem~\ref{THEOREM_OPG} holds.  $\hfill\blacksquare$

\emph{Proof of Lemma~\ref{LTI_IMP}}: It is well known that any strictly proper rational transfer function $H(s)$ can be represented by a triple $(\Pi_c, \upeta_c, \upalpha_c)$ in the following canonical form:
$$H(s)=\frac{b_{1}s^{m-1}+\ldots +b_{m-1}s+b_m}{s^m+a_1s^{m-1}+\ldots+a_{m-1}s+a_m},\;\;\Pi_c=\left[\! \begin{array}{c:c}
       -a_1 & \\
       -a_2 & I_{m-1}\\
       \vdots & \\\cdashline{2-2}
       -a_m & 0
\end{array}\! \right]\!,\;\upeta_c=\left[\!\! \begin{array}{c}
       b_1\\b_2\\\vdots\\b_m
\end{array}\!\! \right]\!,\;\upalpha_c=\left[\! \begin{array}{c}
       1\\0\\\vdots\\0
\end{array}\! \right].$$
If, in addition, $H(s)$ has a unit DC gain, i.e.,  $a_m=b_m\neq 0$, then there always exists a similarity transformation $\mathrm{T}$ such that $H(s)=\upalpha^\top(sI-\Pi)^{-1}\upeta$, where  $(\Pi, \upeta, \upalpha)=(\mathrm{T}^{-1}\Pi_c\mathrm{T}, \mathrm{T}^{-1}\upeta_c, \mathrm{T}^\top\upalpha_c)$, ${\upeta}=-\Pi\mathds{1}_m$, and $\upalpha=[1,0, \ldots, 0]^\top$. An example of such a transformation matrix is a unit lower triangular matrix $\mathrm{T}=[t_{ij}]$, where $t_{ii}=1$, $\forall i$, and $t_{i1}=a_{i-1}-b_{i-1}-1$, for $i=2, \ldots, m$, and all other elements equal to zero. The expression for the step response follows from the properties of matrix exponential functions, that for any invertible square matrix $\Pi$, we have $\int_0^t\text{exp}(\Pi\tau)d\tau=(\text{exp}(\Pi\tau)-I)\Pi^{-1}$.  $\hfill\blacksquare$

\emph{Proof of Lemma~\ref{uniform_conv}}: One approach to construct a global approximation to a function defined in $[0, \infty)$ is expansion in Laguerre polynomials. The sequence $\{\phi_k(t)\}$, where $\phi_k(t)=\sqrt{2\beta}\, e^{-\beta t} L_k(2\beta t)$, $\beta>0$, $k=0, 1, \ldots$, forms a \emph{complete orthonormal set} in $[0, \infty)$, where $L_k(x)=\sum_{i=0}^k((-1)^i/i!)\binom{k}{i}x^i$ is the classical Laguerre polynomial of degree $k$. Then, we have $\int_0^\infty \phi_n^2(t)dt=1$ and $\int_0^\infty \phi_n(t)\phi_m(t)dt=0$, $\forall n,m$, $n\neq m$   \cite[\S 4]{Sansone91}. Since $\{\phi_k(t)\}$ forms a complete orthonormal set in $[0, \infty)$, then any piecewise continuous square-integrable function in $[0, \infty)$ can be approximated  arbitrarily well \emph{in the mean} by a linear combination of $\phi_k(t)$'s \cite{Courant91}. Since $f(t)$ is assumed to be continuous, bounded, and absolutely integrable, then it is square-integrable. Hence,  $h_n(t)=\sum_{k=0}^n a_k \phi_k(t)$, where $a_k=\int_0^\infty f(t)\phi_k(t)dt$, converges in the mean to $f(t)$, i.e.,  $\int_0^\infty |f(t)-h_n(t)|^2dt$ tends to zero, as $n\rightarrow\infty$. The Laplace transform of $\phi_k(t)$ is  $\Phi_k(s)=\sqrt{2\beta}\,(s-\beta)^k/(s+\beta)^{k+1}$, which is a rational function of $s$ and is analytic in $\text{Re}[s]>-\beta$. Then, $H_n(s)=\mathcal{L}[h_n(t)]=\sum_{k=0}^n a_k \Phi_k(s)$ is a stable rational transfer function of order $n$. For the uniform convergence, i.e.,  $\sup_t|f(t)-h_n(t)|\rightarrow 0$ as $n\rightarrow\infty$, however, $f(t)$ must satisfy some smoothness properties. The criteria for the uniform convergence of the series of Laguerre polynomials are given in \cite[\S 4.9]{Sansone91}. According to these criteria, the differentiability of $f(t)$ and the square integrability of its time derivative  ensures the uniform convergence of $h_n(t)$ to $f(t)$, as $n\rightarrow\infty$. It should be noted that, in order to establish uniform global approximation by a rational transfer function, we considered expansion in Laguerre polynomials. This property can be established by using other approaches with different series of polynomials (e.g., mapping $[0, \infty)$ onto a finite interval and using  polynomials defined over finite intervals such as Legendre, Chebyshev, and trigonometric polynomials).    $\hfill\blacksquare$

\emph{Proof of Proposition~\ref{PROP2}}: The existence of a finite $n_0$, such that $h_n(t)\geq 0$, $\forall t\in[0, t_\epsilon]$, $\forall n\geq n_0$, has been established in Proposition~\ref{PROP1}. The filtered version of $h_n(t)$ can be written as $\bar{h}_{n+1}(t)=(p_0/z_0)(h_n(t)+(z_0-p_0)e^{-p_0t}\text{u}_0(t)*h_n(t))$, where $\text{u}_0(t)$ is the unit step function and $*$ denotes the convolution operator. Obviously, for any $z_0\geq p_0>0$, $\bar{h}_{n+1}(t)\geq 0$, $\forall t\in[0, t_\epsilon]$, because $h_n(t)\geq 0$, for $t\in[0, t_\epsilon]$. In order to prove the assertion of the proposition, it suffices to show that, there exist $p_0$ and a finite $\bar n\geq n_0$, such that $e^{-p_0t}\text{u}_0(t)*h_n(t)\geq 0$, $\forall t>t_\epsilon$, $\forall n\geq \bar{n}$. This is because, in this case, from the above expression for $\bar{h}_{n+1}(t)$, by choosing $z_0$ sufficiently away from $p_0$, $\bar{h}_{n+1}(t)$ can be made non-negative $\forall t\geq 0$. Since $H_n(s)$ is stable, there exists $\alpha_0, \zeta_n>0$, such that $h_n(t)\geq -\zeta_n e^{-\alpha_0(t-t_\epsilon)}$, $\forall t>t_\epsilon$, where $\alpha_0$ is the decay rate of the envelope of the tail of $h_n(t)$ (determined by the real part of the dominant pole of $H_n(s)$). We have $e^{-p_0t}\text{u}_0(t)*h_n(t)=e^{-p_0 t}\int_0^t e^{p_0\tau}h_n(\tau)d\tau$, where $\int_0^t e^{p_0\tau}h_n(\tau)d\tau\geq \int_0^{t_\epsilon}h_n(\tau)d\tau - \zeta_ne^{\alpha_0 t_\epsilon}\int_{t_\epsilon}^t e^{-(\alpha_0-p_0)\tau}d\tau=\int_0^{t_\epsilon}h_n(\tau)d\tau - \zeta_n(e^{p_0 t_\epsilon}-e^{\alpha_0t_\epsilon}e^{-(\alpha_0-p_0)t})/(\alpha_0-p_0)\triangleq g(t)$. Then, if $0<p_0<\alpha_0$,  $\inf_{t>t_\epsilon} g(t) = \int_0^{t_\epsilon}h_n(\tau)d\tau - \zeta_ne^{p_0t_\epsilon}/(\alpha_0-p_0)$. The uniform convergence property of $\{
h_n(t)\}$ implies that $\zeta_n$ can be made arbitrary small by increasing $n$. Therefore, there exist $p_0\in(0, \alpha_0)$ and a finite integer $\bar{n}$, such that $\inf_{t>t_\epsilon} g(t)\geq 0$; hence $e^{-p_0t}\text{u}_0(t)*h_n(t)\geq 0$, $\forall t>t_\epsilon$, $\forall n\geq\bar{n}$. Using the triangle inequality, we have $\left|\|f-h_n\|_p-\|f - \bar{h}_{n+1}\|_p\right|\leq \|h_n-\bar{h}_{n+1}\|_p$, where $h_n(t)-\bar{h}_{n+1}(t)=(1-p_0/z_0)h_n(t)-(1-p_0/z_0)(p_0e^{-p_0t}\text{u}_0(t))*h_n(t)$. Then, $\|h_n-\bar{h}_{n+1}\|_p\leq (1-p_0/z_0)(\|h_n\|_p + \|p_0 e^{-p_0t}\text{u}_0(t)\|_1\|h_n\|_p)\leq 2(1-p_0/z_0)\|h_n\|_p$, where $\|h_n\|_p$ is bounded $\forall p\in[1, \infty]$, as $H_n(s)$ is stable. Moreover, since $\bar{H}_{n+1}(0)=H_n(0)W(0)=1$, then $\bar{h}_{n+1}(t)$ integrates to one, and hence is a valid pdf.  $\hfill\blacksquare$


\end{document}